\documentclass[aps,pre,showpacs,amsmath]{revtex4}
\usepackage{graphicx}% Include figure files
\def\eps{\varepsilon}
\def\Dm{\widetilde{\cal D}_{\mu}}
\def\E{\overline{\cal E}}

\def\st#1{\left[#1\right]}

\newcommand{\bx}{{\bf x}}

\newcommand{\be}{\begin{equation}}
\newcommand{\ee}{\end{equation}}
\newcommand{\Fa}{\vphantom{\vdots}}
\newcommand{\berr}{\be\begin{array}{c}}
\newcommand{\berrl}{\be\begin{array}{l}}
\newcommand{\eerr}{\Fa\end{array}\ee}
\newcommand{\eerrl}{\end{array}\ee}

\begin{document}
\title {An improved $\eps$ expansion for three-dimensional turbulence:
two-loop renormalization near two dimensions}
\author {L.Ts.~Adzhemyan$^1$, J.~Honkonen$^2$,
M.V.~Kompaniets$^1$ and A.N.~Vasil'ev$^1$}

\affiliation{$^1$ Department of Theoretical Physics, St.~Petersburg
University, Uljanovskaja 1, St.~Petersburg,
Petrodvorets, 198504 Russia, $^2$ Theoretical Physics~Division,
Department~of~Physical Sciences, P.O.~Box~64,FI-00014
University~of~Helsinki, Finland}

\begin{abstract}
An improved $\eps$ expansion in the $d$-dimensional ($d > 2$)
stochastic theory of turbulence is constructed at two-loop order
which incorporates the effect of pole singularities
at $d \rightarrow 2$ in coefficients of the $\eps$ expansion of universal quantities.
For a proper account of the effect of these singularities two different approaches to the
renormalization of the powerlike correlation function of the random force are analyzed near two
dimensions. By direct calculation it is shown that
the approach based on the mere renormalization of the nonlocal correlation function
leads to contradictions at two-loop order. On the other hand, a two-loop
calculation in the renormalization scheme
with the addition to the force correlation function of a local term to be renormalized instead of the
nonlocal one yields consistent results
in accordance with the UV renormalization theory. The latter renormalization
prescription is used for the two-loop renormalization-group analysis amended
with partial resummation of the pole singularities near two dimensions leading to a significant
improvement of the agreement with experimental results for the Kolmogorov constant.
\end{abstract}


\date{\today}

\maketitle

\section{\label{sec:intro}Introduction}

The renormalization-group (RG) method in the theory of turbulence allows to calculate various
physical quantities --- critical exponents and universal amplitude ratios --- in the form of an
expansion in a small parameter $\eps$. The real value of this parameter is not small, however,
therefore justified doubts arise whether this method is of any use for acceptable numerical
estimates of the quantities studied. Until recently, practical calculations were carried out only
in the simplest (one-loop) approximation and therefore it was not possible to assess how the
next-to-leading terms of the expansion actually compare with the leading order at the real value
of the parameter $\eps$. In Refs.
\cite{Adzhemyan03a} and \cite{slovac}
this problem was analyzed on the example of
calculation of the skewness factor and the Kolmogorov constant in the inertial range. The calculation
showed that the relative part of the two-loop correction is indeed large, of the order of 100 \% in
the real space dimension $d=3$. This contribution, however, rapidly decreases with the growth
of $d$: already for $d=5$ it yields only 30 \% and in the limit $d\to\infty$ decreases to 10 \%.
On the contrary, when the space dimension decreases from $d=3$ to $d=2$ a drastic growth of the
correction term is observed.

Analysis of the dependence of the coefficients of the
$\eps$ expansion on the space dimension has revealed that this property is connected with the
divergence of some graphs in the limit $d\to 2$, and the singularities in $ d-2\equiv 2 \Delta$
accumulate with the order of the perturbation expansion. Contributions of these graphs turned
out to give rise to the large value of the correction term also at $d=3$. Thus, satisfactory
quantitative results may be expected only after summing, at least approximately, the contributions
of the most singular graphs at all orders of the $\eps$ expansion. Such a summation has been carried out
in Ref. \cite{Adzhemyan03b} with the use of an additional renormalization and double expansion in
$\varepsilon$ and $\Delta$ and with the result of a significant relative reduction of the
correction term and improvement of the agreement with experiment. The calculation in Ref. \cite{Adzhemyan03b}
has been carried out in the two-loop approximation for both the usual $\eps$ expansion and the
double ($\eps$, $\Delta$) expansion. These expansions were used as complementary to each other to
arrive at the final result -- an approach distinguishing Ref. \cite{Adzhemyan03b} from
Refs. \cite{Ronis,Nalimov}, in which the one-loop calculation in the ($\eps$, $\Delta$) expansion
was carried out.

In Ref. \cite{Ronis}, where the idea of the double expansion was first applied to the stochastic
Navier-Stokes problem, the method used to eliminate the additional
divergences was the natural at first sight multiplicative renormalization
of the nonlocal correlation function of the random force
(we shall further call this the nonlocal scheme). In Ref. \cite{Nalimov} a different
scheme of renormalization was adopted based on the general statement of the theory of
UV renormalization that the counterterms are local.
In the one-loop approximation, to which the authors of Refs. \cite{Ronis,Nalimov} restricted
themselves, it is possible to remove the divergences of the graphs both in the nonlocal
renormalization scheme of Ref. \cite{Ronis} and in the local scheme of Ref. \cite{Nalimov}, so that
at this level both approaches seem equally acceptable. This, however, is not so already in the
next two-loop approximation.
In this paper we show that the renormalization scheme of Ref. \cite{Ronis} is not consistent at
two-loop order, because within it the two-loop contributions to the renormalization constants
acquire an inacceptable dependence on the external wave vectors, whereas in the scheme of local
renormalization \cite{Nalimov} this does not occur. The appearance of the latter scheme is more
complicated and the reader unfamiliar with the delicacies of the RG approach might have doubts
in the necessity to use it, all the more so because the argumentation in its favor requires
a discussion of the so called $\Lambda$ renormalization (similar to the shift of the
critical temperature in the theory of critical phenomena).
This paper is devoted to a detailed
analysis of these issues. It also contains a technical account of the method which
allowed to obtain the
two-loop results announced in Rapid Communication \cite{Adzhemyan03b}.

This paper is organized as follows. In Sec. \ref{sec:d>2} we remind
basic features of the
(field-theoretic) renormalization procedure and the
subsequent asymptotic analysis in the two-loop approximation well above the problematic two dimensions.
Sec. \ref{sec:ARonis} is devoted to a detailed argument showing why the multiplicative
nonlocal renormalization fails at the two-loop order of the double expansion.
In Sec. \ref{sec:local} the consistency of  the local two-charge renormalization
scheme is demonstrated by the results of the two-loop calculation in space dimensions
$d\le 2$ in which the technically simplest combined scheme of analytic and dimensional
renormalization is unconditionally valid. The fairly technical issue of the
possibility of analytic continuation
of the results obtained for $d\le 2$ to space dimensions above two is demonstrated
in the Appendix.
Renormalization-group equations are set up in Sec. \ref{sec:RGE} with
the subsequent two-loop solution for asymptotic analysis in the inertial range.
Details of the method of calculation of universal quantities
in the improved $\eps$ expansion are exposed in Sec. \ref{sec:Q}. Sec. \ref{sec:conclusion}
contains discussion of the results and concluding remarks.

\section{\label{sec:d>2}Renormalization of the model in a fixed space dimension $d>2$}

The statistical model of the developed homogeneous isotropic turbulence of incompressible
fluid is based on the stochastic Navier-Stokes equation
\begin{equation}
\nabla _t\varphi _i=\nu _0\partial^{2} \varphi _i-\partial _i
{\cal P}+f_i , \qquad \nabla _t\equiv \partial _t+(\varphi
\partial)  . \label{1.1}
\end{equation}
Here,
$\varphi _i(t,\bx)$ is the divergenceless velocity field,
${\cal P}(t,\bx)$ and $f_i(t,\bx)$ the pressure and the transverse
random force per unit mass, respectively,
$\nu _0$ the kinematic viscosity. For the random force
$f$ a Gaussian distribution is assumed with zero mean and the
correlation function
\begin{equation}
\big\langle f_i(t,{\bf x})f_j(t',{\bf x'})\big\rangle \equiv
D_{ij}(t,{\bf x};t',{\bf x'})
= \frac{\delta (t-t')}{(2\pi)^{d}}\,
\int d{\bf k}\, P_{ij}({\bf k})\, d_f(k)\, \exp \big[{\rm i}{\bf
k} \left({\bf x}-{\bf x}'\right)\big] , \label{1.2}
\end{equation}
where
$P_{ij}({\bf k}) =\delta _{ij}  - k_i k_j / k^2$
is the transverse projection operator,
$d$ the dimension
of the coordinate space.
For the function
$d_f(k)$ the following powerlike form is adopted in the RG approach:
\be
d_f(k)=D_0 k^{4-d-2\eps}. \label{nakach}
\ee
The quantity
$\eps > 0$ in Eq. (\ref{nakach}) plays the role of a formal expansion
parameter. The value corresponding to the physical model is
$\eps=2$, because for
$\varepsilon\rightarrow 2$, $D_0\sim (2-\varepsilon)$
we arrive at
$d_F(k)\sim \delta(\bf k)$ which
corresponds to energy injection by infinitely large eddies.

The stochastic problem
(\ref{1.1}) and (\ref{1.2})
is equivalent to a quantum-field-theoretic model with a
doubled set of transverse vector fields
$\Phi\equiv\{\varphi,\varphi'\}$
and the action \cite{Dominicis79}
\begin{equation}
S(\Phi )=\varphi 'D\varphi '/2+\varphi '[-\partial _t\varphi +\nu
_0\partial^{2} \varphi -(\varphi \partial )\varphi ] ,
\label{action}
\end{equation}
where
$D$
is the correlation function of the random force (\ref{1.2}),
and the necessary integrals over
($t$, ${\bf x}$)
and sums over vector indices are implied.
Action
(\ref{action})
gives rise to the standard diagrammatic technique
with the bare propagators whose
($t$, ${\bf k}$) representation is of the form
\begin{align}
\label{lines}
\langle \varphi(t) \varphi '(t')\rangle
_0&= \theta(t-t') \exp \left[-\nu_0 k^{2} (t-t') \right],\nonumber\\
\langle\varphi'\varphi'\rangle _0&=0, \\
\langle \varphi \varphi\rangle _0&={d_f(k)\over 2\nu_0 k^{2}} \exp \left[-\nu_0 k^{2} |t-t'| \right]\,,\nonumber
\end{align}
where the common factor
$P_{ij}({\bf k})$ has been omitted for simplicity. The interaction in Eq. (\ref{action})
brings about the three-point vertex
$-\varphi'(\varphi\partial)\varphi=\varphi'_iV_{ijs}\varphi_j\varphi_s/2$
with the vertex factor
$V_{ijs}={\rm i}(k_j\delta _{is}+k_s\delta _{ij})$,
where
${\bf k}$ is the wave vector of the field
$\varphi'$. The expansion parameter of the perturbation theory is
the coupling constant
$g_0\equiv D_0/\nu_0^3$.

Model
(\ref{action}) is logarithmic (i.e. the coupling constant $g_0$ is
dimensionless) at $\eps=0$. In the analytic renormalization scheme
adopted here the UV divergences have the form
of the poles in $\eps$ in the correlation functions of the field
$\Phi\equiv\{\varphi,\varphi'\}$. Dimensional analysis (power
counting) shows that for $d>2$ superficial UV divergences can be present only in the
one-particle-irreducible (1PI)
functions $\Gamma_{\varphi'\varphi}$ and
$\Gamma_{\varphi'\varphi\varphi}$. These
divergences may be removed by
counterterms of the form
\begin{equation}
\label{counterterms}
\varphi'\partial^{2}\varphi\,,\qquad
\varphi'\partial_t\varphi\,,\qquad
\varphi'(\varphi\partial)\varphi
\end{equation}
in the action. Due to symmetry reasons, however, in model
(\ref{action}) only one counterterm of all allowed
by the dimensional analysis is actually generated. First, the spatial derivative acting on the
field $\varphi$ in the interaction term of action
(\ref{action}) can be transferred to the field $\varphi'$ with the use of
integration by parts. This means that the counterterms to the
1PI functions must contain at least one spatial derivative, so that
the structure
$\varphi'\partial_t\varphi$ cannot possibly appear.
Second, from the Galilean symmetry of
action (\ref{action}) it follows that the last two structures
of Eq. (\ref{counterterms}) can
be brought about as counterterms only in the invariant combination
$\varphi'\nabla_t\varphi$ with the Lagrangian derivative
$\nabla_t=\partial_t+(\varphi\partial)$ from Eq. (\ref{1.1}). This
excludes also the structure $\varphi'(\varphi\partial)\varphi$.
Thus, in the generic case we are left with a single counterterm
of the form $\varphi'\partial^{2}\varphi$.
In the special case $d=2$, however, a new UV divergence appears in the
1PI function $\Gamma_{\varphi'\varphi'}$.

Consider the renormalization of model
(\ref{action})
in the two-loop approximation in
$d>2$. In this case the only counterterm required is
$\varphi'\partial^{2}\varphi$ which is generated by multiplicative
renormalization of the viscosity in the corresponding term of action
(\ref{action}). We shall use the scheme of minimal subtractions (MS)
in which the renormalization constants are determined by the relations
\begin{align}
\nu_0&=\nu Z_{\nu},&D_{0}&= g_{0}\nu_0^{3} = g\mu^{2\eps}
\nu^{3}, \nonumber\\
g_{0}&=g\mu^{2\eps}Z_{g}, &Z_{g}&=Z_{\nu}^{-3}. \label{18}
\end{align}
Here,
$\mu$ is the scale-setting parameter (the reference mass) in the MS scheme,
$\nu$ is the renormalized viscosity and $g$ the dimensionless renormalized charge.
The only independent renormalization constant in Eq.
(\ref{18}) is that of the viscosity
$Z_{\nu}$. The amplitude of the correlation function of the random force
$D_{0}$ is not renormalized, because no counterterm of the form
$\varphi' \varphi'$ in action (\ref{action}) is necessary. This leads to the relation
between the renormalization constants of the charge and viscosity indicated in Eq. (\ref{18}).

In the MS scheme the renormalization constants are constructed
as Laurent series in $\eps$ of the form
''1 + $\sum_{n\ge 1}{ a_n\eps^{-n}}$''.
In particular,
\begin{equation}
Z_{\nu}=1+u\,\frac{a_{11}}{\varepsilon}+u^2\,\left(\frac{a_{22}}{\varepsilon^2}+\frac{a_{21}}{\varepsilon}\right)+...
=1+\sum _{n=1}^{\infty }u^n \sum _{k=1}^{n}a_{nk}\eps ^{-k\,},
\label{1.30}
\end{equation}
where
\begin{equation}
u\equiv {g\bar S_{d}\over 32}\,,\quad
 \bar S_d \equiv {S_d \over (2\pi)^{d}},\quad S_d \equiv {2\pi^{d/2}\over \Gamma (d/2)}\,,\label{surface}
\end{equation}
and the coefficients $a_{nk}$ depend only on $d$. Here $S_d $
is the surface area of the unit sphere in $d$-dimensional space
and $\Gamma$ is Euler's Gamma function.

We shall determine the constant $Z_{\nu}$ from the requirement
that the 1PI correlation function
$\Gamma_{\varphi'\varphi}$ at zero frequency
($\omega=0$) is UV finite, i.e.  finite at $\eps\to 0$ when
expressed as a function of the renormalized variables $\nu$ and $g$ determined
by relations (\ref{18}). With respect to vector indices the function
$\Gamma_{\varphi'\varphi}$ is proportional to the transverse projector
$P_{ij}(\bf p)$, where $\bf p$ is the external wave vector.
In the following we shall deal with the scalar coefficient of this
projector obtained by the contraction of the indices
$i$ and $j$ and division by
$\text{Tr}\, P=d-1$. In terms of the bare parameters
$\nu_0$ and $D_0=g_0\nu_0^3$ this scalar coefficient at $\omega=0$
assumes the form $-\nu_0 p^2+$ sum of contributions of the
$n$-loop graphs, each of which contains $n$ pieces of $\langle \varphi \varphi \rangle _0$
lines (\ref{lines}) and, correspondingly, the factor $D_0^n$. Thus,
in view of dimensional arguments
\begin{eqnarray}
\frac{\text{Tr}\,\Gamma_{\varphi'\varphi}\left|_{\omega=0}\right.}{d-1}
= \nu_0
p^2\left[-1+\sum_{n=1}^{\infty}\left(\frac{D_0\bar S_{d}}{32\nu_0^3
p^{2\varepsilon}}\right)^n \gamma^{(n)}_{\varphi'\varphi}\right]
\label{ti}
\end{eqnarray}
with dimensionless coefficients
$\gamma^{(n)}_{\varphi'\varphi}$
which only depend on
$d$ and $\varepsilon$. The factors 32  and $\bar S_{d}$
in Eq. (\ref{ti}) have been introduced for convenience.
To obtain the renormalized function $\Gamma_{\varphi'\varphi}$
the parameters
$D_0$ and $\nu_0$
in Eq. (\ref{ti}) have to be expressed in terms of
$\nu$, $g$ and $\mu$ according to definitions
(\ref{18}), which leaves the coefficients
$\gamma^{(n)}_{\varphi'\varphi}$ intact.
It is convenient to divide the result by
$\nu p^2$ to arrive at the dimensionless quantity
\begin{equation}
\frac{\text{Tr}\,\Gamma_{\varphi'\varphi}|_{ \omega=0}} {\nu p^2(d-1)}=-Z_\nu
+u s^{2\varepsilon}\,
\gamma_{\varphi'\varphi}^{(1)}\,Z_\nu^{-2}+(u
s^{2\varepsilon})^2\,\gamma_{\varphi'\varphi}^{(2)}\,Z_\nu^{-5}+...
\label{Ratio}
\end{equation}
with $u$ from Eq. (\ref{surface}) and $s\equiv\mu/p$.

The renormalization constant $Z_\nu$ is determined from the
condition of cancellation of the poles in $\varepsilon$ in relation (\ref{Ratio}).
In the coefficient $\gamma_{\varphi'\varphi}^{(1)}$ there is a simple pole $\sim 1/\varepsilon$,
whereas $\gamma_{\varphi'\varphi}^{(2)}$ contains poles $\sim1/\varepsilon$ and
$\sim 1/\varepsilon^2$ etc. For the two-loop calculation of $Z_\nu$
the following contributions are needed:
\begin{align}
\gamma_{\varphi'\varphi}^{(1)}&=\frac{A}{\varepsilon}+B+...,\label{AB}\\
\gamma_{\varphi'\varphi}^{(2)}&=\frac{C}{\varepsilon^2}+\frac{D}{\varepsilon}+...
, \label{CD}
\end{align}
where the ellipsis stands for irrelevant corrections $O(\varepsilon)$
in $\gamma_{\varphi'\varphi}^{(1)}$ and $O(1)$ in
$\gamma_{\varphi'\varphi}^{(2)}$.

Denoting the contribution of the order $u^n\sim g^n$  to the renormalization
constant (\ref{1.30}) by $Z_\nu^{(n)}$, from the condition of cancellation
of the divergences (poles in $\varepsilon$) in Eq. (\ref{Ratio}) we infer
\begin{align}
Z_\nu^{(1)}&=\mathcal{L}_\eps\,
\left[us^{2\varepsilon}\gamma_{\varphi'\varphi}^{(1)}\right],
 \label{1z}\\
Z_\nu^{(2)}&=\mathcal{L}_\eps
\,\left[u^2s^{4\varepsilon}\gamma_{\varphi'\varphi}^{(2)}-
2Z_\nu^{(1)}us^{2\varepsilon}\gamma_{\varphi'\varphi}^{(1)}\right],
 \label{2z}
\end{align}
where
$\mathcal{L}_\eps$ stands for the operation of
extraction of the UV-divergent part, which here consists of poles in $\varepsilon$.

When relation (\ref{AB}) is substituted in  Eq. (\ref{1z}) the
UV-finite term $B$ does not contribute and the coefficient
$s^{2\epsilon}=1+2\varepsilon\log s +...$
may be replaced by the unity. As a result we obtain
\begin{equation}
Z_\nu^{(1)}=\frac{uA}{\varepsilon}.
 \label{3z}
\end{equation}
Substituting this expression together with relations (\ref{AB}) and (\ref{CD}) in Eq. (\ref{2z}) we find
\begin{equation}
Z_\nu^{(2)}=
\mathcal{L}_\eps\,\left[u^2\,s^{4\varepsilon}\left(\frac{C}
{\varepsilon^2}+\frac{D}{\varepsilon}\right)
%\right.\\\left.
-2u^2\,s^{2\varepsilon}\frac{A}{\varepsilon}\left(\frac{A}
{\varepsilon}+B\right)\right].
 \label{30z}
\end{equation}
In the terms $\sim1/\varepsilon$ we may replace
$s^{n\varepsilon}\rightarrow 1$, whereas in contributions
$\sim1/\varepsilon^2$  also the second term in the expansion
$s^{n\varepsilon}=1+n\varepsilon \log s + ...$ must be
retained which gives rise to a contribution
of the form $\varepsilon^{-1}\log s =\varepsilon^{-1}\log
(\mu/p)$ in $Z_\nu^{(2)}$. The presence of such a term in $Z_\nu$
is inacceptable, because renormalization constants must not contain
any wave-number dependence by their very definition. The condition
of vanishing of the term $\sim\varepsilon^{-1}\log s$ in (\ref{30z}) is
\begin{eqnarray}
C=A^2
\label{usl}
\end{eqnarray}
for the coefficients of relations (\ref{AB}) and (\ref{CD}).

The recent two-loop calculation \cite{Adzhemyan03a} confirms
that relation (\ref{usl}) holds. Substituting  it in Eq.
(\ref{30z}) we obtain
\begin{eqnarray}
Z_\nu^{(2)}=u^2\,\left[-\frac{A^2}{\varepsilon^2}+\frac{D-2AB}{\varepsilon}\right].
 \label{4z}
\end{eqnarray}
The one-loop coefficient $A$ in Eqs. (\ref{3z}), (\ref{30z}) and (\ref{4z}) has been known
for quite a while:
\[
A=-\frac{4(d-1)}{d+2}.
% \label{7z}
\]
For the nontrivial next-to-leading coefficients $D$ and $B$ in Eq. (\ref{4z}) integral
representations readily calculable for
any given $d$ have been obtained in Ref. \cite{Adzhemyan03a}.

That condition
(\ref{usl}) holds thus imposing on $Z_\nu$ cancellation of
the contributions $\sim\log s$
is not a coincidence, but a consequence of general principles of the theory
of UV renormalization. The most important of them is the requirement that all counterterms
must be local in space (i.e. polynomial in wave vectors). In model
(\ref{action}) this is so, because the counterterm giving rise to the renormalization
of the parameter $\nu_0$ has the form of
$\nu p^2$ multiplied by a wave-number-independent coefficient, i.e.
a polynomial function in $\bf p$. Therefore in this model all consequences of the general
conjectures of the theory of UV renormalization must hold, in particular,
independence of the renormalization constants of wave numbers to all orders in the perturbation
theory as well as the critical scaling due to the RG equations with the $\eps$-dependent
critical dimensions of the velocity field $\varphi$
and the frequency $\omega$ (more details in Sec.\ref{sec:RGE}):
\begin{eqnarray}
\Delta_\varphi=1-2\varepsilon/3,\qquad
\Delta_\omega=2-2\varepsilon/3.
 \label{razm}
\end{eqnarray}
These are exact relations without any corrections of higher order in
$\varepsilon$. They are a consequence of connection (\ref{18}) between
the renormalization constants
$Z_g$ and $Z_\nu$ which, in turn, follows from the absence of renormalization
of the nonlocal contribution with the correlation function of the random force
in action (\ref{action}).
At the real value
$\varepsilon=2$ quantities (\ref{razm}) assume the Kolmogorov values
\begin{eqnarray}
\Delta_\varphi=-1/3,\qquad \Delta_\omega=2/3.
 \label{kolm}
\end{eqnarray}
Condition (\ref{usl}) ensuring independence of the renormalization
constant of the wave number in the MS scheme may appear in a
different form in other renormalization schemes. We will
illustrate this point on the example of the scheme with the
''normalization point'' (NP). In this approach the renormalization
constant $Z_\nu$ is calculated from the normalization condition
for the 1PI Green function
\begin{eqnarray}
\frac{\text{Tr}\,\Gamma_{\varphi'\varphi}|_{ \omega=0}} {\nu
p^2(d-1)}\Biggl|_{p=\mu}
 = -1
\label{point}
\end{eqnarray}
in contrast to the cancellation of poles in $\varepsilon$ in
expression (\ref{Ratio}) in the MS scheme. Then instead of Eqs.
(\ref{1z}) and (\ref{2z}) we obtain
\begin{align}
\label{point1}
Z_\nu^{(1)}&= \, u\gamma_{\varphi'\varphi}^{(1)}\,,\nonumber\\
Z_\nu^{(2)}&= \,u^2\gamma_{\varphi'\varphi}^{(2)}-
2Z_\nu^{(1)}u\gamma_{\varphi'\varphi}^{(1)}\,
%\\&
=\,u^2\Big[\gamma_{\varphi'\varphi}^{(2)}-
2(\gamma_{\varphi'\varphi}^{(1)})^2 \Big],
\end{align}
and after substitution of $Z_\nu$ from Eq. (\ref{point1})  expression
(\ref{Ratio}) assumes the form
\begin{equation}
\label{point2} \frac{\text{Tr}\,\Gamma_{\varphi'\varphi}|_{ \omega=0}} {\nu
p^2(d-1)}= -1+u \,
\gamma_{\varphi'\varphi}^{(1)}\,(s^{2\varepsilon}-1)
+u^2\Big[
\,\gamma_{\varphi'\varphi}^{(2)}(s^{4\varepsilon}-1)
-2(\gamma_{\varphi'\varphi}^{(1)})^2(s^{2\varepsilon}-1)\Big]
 + O(u^3).
\end{equation}
In the NP scheme the renormalization constant  $Z_\nu$ does not
depend on  $s=\mu/p$ due to the very definition, but
cancellation of poles in $\varepsilon$ is not obvious in Eq.
(\ref{Ratio}). In the two-loop approximation (\ref{point2}) with
account of expressions (\ref{AB}) and (\ref{CD}) these poles appear in the
form $\sim u^2\varepsilon^{-1}\log s$ in several contributions,
and the condition of their mutual cancellation is the same
relation (\ref{usl}) which ensured the cancellation of the ''bad''
contributions $\sim u^2 \varepsilon^{-1}\log s$ in $Z_\nu$ in
the MS scheme. As it was previously explained, fulfilment of
condition (\ref{usl}) is guaranteed by general theorems of the
theory of UV renormalization with local counterterms.

The MS and NP schemes differ by a finite renormalization of the
parameters $g$ and $\nu$, therefore all objective physical
quantities, in particular, critical dimensions  (\ref{razm}),
calculated in these schemes coincide.

Critical dimensions
(\ref{razm}) do not depend on $d$ and thus for them the problem of singularities
in the limit $d\rightarrow 2$ mentioned in Sec. \ref{sec:intro} is not relevant.
There are, however, other important physical quantities such as the skewness
factor, Kolmogorov constant, critical dimensions of various composite operators
to which this problem persists. It is important that for these quantities the
problem of anomalous scaling is absent, which cannot be treated in the framework
of the model with massless injection (\ref{nakach}) lacking a dimensional parameter
to account for the external scale of turbulence.

For such quantities, contrary to Eq. (\ref{razm}), the solutions contain full series of the form
\begin{equation}
R(\eps,d)=\sum_{k=0}^\infty R_k(d)\eps^k ,  \label{Reps}
\end{equation}
and the coefficients
$R_k(d)$ in the limit $d\rightarrow 2$
reveal singular behavior of the type
$\sim(d-2)^{-k}\sim\Delta^{-k}$
($2\Delta\equiv d-2$)
giving rise to the growth of the relative part of the correction terms at
$d\rightarrow 2$. The effect of these is fairly discernable also at the
real value $d=3$, hence the natural desire to sum up contributions of the form
$(\varepsilon/\Delta)^k$ at all orders of the
$\varepsilon$ expansion (\ref{Reps}). This may be done with the aid of the
double ($\varepsilon$, $\Delta$) expansion \cite{Ronis, Nalimov}. The idea of such
an ''improved $\varepsilon$ expansion'' with the use of the local
renormalization scheme  \cite{Nalimov} was explained in our Rapid Communication \cite{Adzhemyan03b},
where many important subtleties and details of calculations were, however,
not reflected due to lack of space. In the present paper we give a detailed exposition
and start from the proof of inconsistency of the renormalization scheme proposed in Ref. \cite{Ronis}.

\section{\label{sec:ARonis}Construction of the double ($\eps$, $\Delta$) expansion. Proof of
the inconsistency of the nonlocal renormalization  \cite{Ronis} in the two-loop
approximation}

Model (\ref{action}) is logarithmic (i.e. the bare coupling constant $g_0$ is dimensionless)
at $\varepsilon=0$ in function (\ref{nakach})
in arbitrary space dimension $d$. In a fixed dimension $d>2$ the  value $\varepsilon=2$
corresponds to the ''real problem''. Calculations in the
framework of the
$\varepsilon$ expansion have a rigorous meaning only in the vicinity of
$\varepsilon=0$, whereas continuation of the results to the ''real'' value
$\varepsilon=2$ is always understood as an extrapolation. In the scheme
applicable for $d>2$ reviewed in
Sec. \ref{sec:d>2} this extrapolation corresponds to the continuation
along the vertical ray from the point
$(d,\,\varepsilon=0)$ to the point  $(d,\,\varepsilon=2)$
in the ($d$, $\varepsilon$) plane. The same final point may be reached along
a ray from any starting point
$(d_0\neq d,\,\varepsilon=0)$ at which the model is logarithmic as well.
The extrapolation along the ray starting from the origin
$(d_0=2,\,\varepsilon=0)$ is, however, singled out, because at $d=2$ in model (\ref{action})
an additional UV divergence (absent at $d>2$) occurs in the 1PI function
$\Gamma_{\varphi'\varphi'}$. On such a ray we put
\begin{eqnarray}
d=2+2\Delta,\qquad\Delta/\varepsilon=\zeta=const.
 \label{dz}
\end{eqnarray}
The parameters $\varepsilon$ and  $\Delta$ are considered small of the same order
and their ratio $\Delta/\varepsilon=\zeta$  a fixed constant
[$\zeta=1/4$ in the extrapolation to the point
$(d=3,\,\varepsilon=2)$].

Extraction of contributions of the order
$\varepsilon^m$ with
$\Delta/\varepsilon=const$ corresponds to the account of all contributions
of the form
$\varepsilon^m(\varepsilon/\Delta)^n$ with any $n=0,\,1,\,2...$
and $m+n=k$ in Eq.
(\ref{Reps}). Thus the use of the ($\varepsilon$, $\Delta$) expansion
in such a form is directly related to the problem of the account of
the singularities at $\Delta\rightarrow 0$ pointed out in the discussion of
relation (\ref{Reps}).

\begin{figure}[!]
\includegraphics[width=9cm]{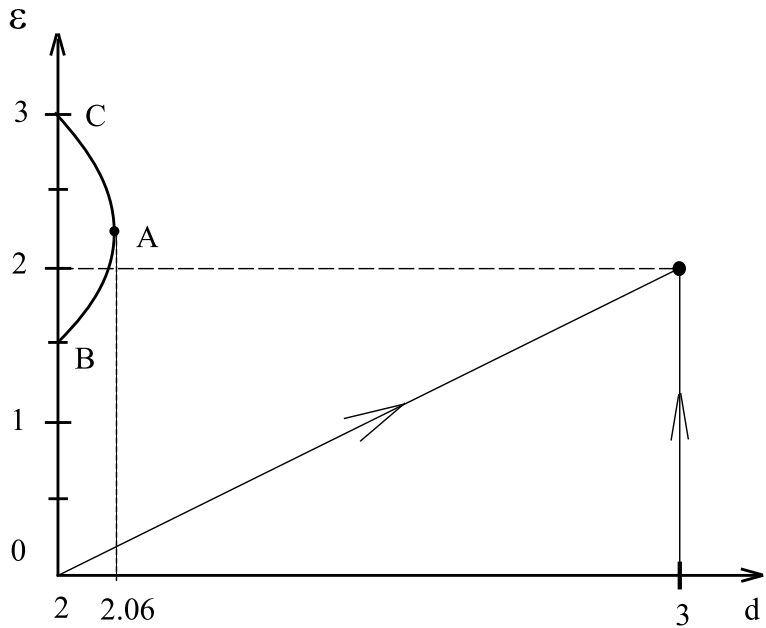}
\vspace{3mm} \caption{\label{frish2}The borderline BAC between the
regions of parameter space $d$, $\eps$ corresponding to direct (to
the right from the curve BAC) and inverse (to the left) energy
cascades.}
\end{figure}

It is worth emphasizing that the very process of extrapolation along a ray
from the starting point
$(d=2,\,\varepsilon=0)$ is inapplicable to description of two-dimensional
turbulence in which the physics is totally different from the three-dimensional
problem due to the appearance of the inverse energy cascade \cite{Legacy}.
In Fig. \ref{frish2} we have plotted the borderline curve BAC between the
direct (normal) and inverse energy cascades obtained in Ref. \cite{Fournier1977}.
The starting point of the extrapolation for the two-dimensional case
$(d=2,\,\varepsilon=0)$ lies in the region of the direct cascade, whereas the final
point $(d=2,\,\varepsilon=2)$ in the region of the inverse cascade. Thus the ray
connecting these points intersects the borderline -- the curve BAC -- so that the
extrapolation becomes impossible. However, the ray connecting the starting point
$(d=2,\,\varepsilon=0)$ and a final point like
$(d=3,\,\varepsilon=2)$ lies completely in the region of the direct cascade, therefore on
such a ray the problem of the change of the cascade pattern does not arise. The rightmost
point of the region of the inverse cascade (point A on Fig. \ref{frish2}) has the coordinate
$d_A\simeq 2.06$ \cite{Fournier1977}. In the preceding discussion of the extrapolation along
the vertical ray from the point
$(d,\,\varepsilon=0)$ to the point $(d,\,\varepsilon=2)$ at $d>2$,
it should have been noted that the condition is not simply $d>2$, but $d>d_A=2.06$.
From the practical point of view this is irrelevant, because we are interested in the
space dimension $d=3$.

The idea of the double
($\varepsilon$, $\Delta$) expansion together with the extrapolation
along the ray
$\Delta\sim\varepsilon$ of relation (\ref{dz}) in the context of the present problem was first
put forward in Ref. \cite{Ronis}. The UV divergences are present not only
in the 1PI function
$\Gamma_{\varphi'\varphi}$ but also in
$\Gamma_{\varphi'\varphi'}$ and appear in the form of poles in the parameters
$\varepsilon$ and $\Delta$ and linear combinations thereof, or, equivalently,
as poles in
$\varepsilon$ with the fixed ratio
$\Delta/\varepsilon\equiv\zeta=const$. To remove the additional divergences
from the graphs of the 1PI function
$\Gamma_{\varphi'\varphi'}$ renormalization of the amplitude
$D_0$ in the nonlocal correlation function of the random force
(\ref{1.2}) and (\ref{nakach}) was used in Ref. \cite{Ronis}, i.e. relations (\ref{18}) between
bare and renormalized parameters were replaced by
\begin{align}
\nu_0&=\nu {Z}_{\nu},  &D_{0}& = g_{0}\nu_0^{3} =
g\mu^{2\eps} \nu^{3}{Z}_D,\nonumber\\
g_{0}&=g\mu^{2\eps}{Z}_{g}, &{Z}_{g}{Z}_{\nu}^{3}&={Z}_D
\label{19}
\end{align}
with a new renormalization constant
${Z}_D$ which does not have an analog in Eq. (\ref{18}).

It should be noted that the introduction of the additional constant
${Z}_D$ breaks the last connection in Eq.
(\ref{18}) and its consequences
(\ref{razm}). Therefore, the author of Ref. \cite{Ronis}
has put forward the conjecture that in the scheme of the double
($\varepsilon$, $\Delta$) expansion at the real value
$\varepsilon=2$ the velocity field
$\varphi$ and the frequency
$\omega$ have dimensions with values different from the Kolmogorov
values (\ref{kolm}). This is, of course, true, if renormalization
relations (\ref{19}) are used. We shall further show, however, that
the renormalization scheme of Ref. \cite{Ronis} with relations
(\ref{19}) is not internally consistent. This
is not obvious in the one-loop approximation,
to which the author of Ref. \cite{Ronis} restricted himself, but becomes
apparent already in the next two-loop approximation. In Ref. \cite{Nalimov}
another scheme of construction of the double
($\varepsilon$, $\Delta$) expansion was put forward in which the last
equality in Eq. (\ref{18}) together with its consequences (\ref{razm}) and (\ref{kolm})
are preserved. We shall deal with this approach in Sec. \ref{sec:local}.

The main goal of this section is to prove that the scheme
of multiplicative renormalization (\ref{19}) contains intrinsic
contradictions. To this end, consider representations similar to
(\ref{Ratio}) for the 1PI functions
 $\Gamma_{\varphi'\varphi}$ and
$\Gamma_{\varphi'\varphi'}$. According to Eq.
(\ref{19}) the amplitude $D_0$  in Eq. (\ref{ti}) now acquires the additional
factor ${Z}_D$, therefore instead of relation
(\ref{Ratio}) we now obtain
\begin{equation}
\frac{\text{Tr}\,\Gamma_{\varphi'\varphi}\left|_{ \omega=0}\right.}{\nu p^2(d-1)}=-{Z}_\nu
+us^{2\varepsilon}\,
\gamma_{\varphi'\varphi}^{(1)}\,{Z}_\nu^{-2}{Z}_D
+(u
s^{2\varepsilon})^2\,\gamma_{\varphi'\varphi}^{(2)}\,{Z}_\nu^{-5}{Z}_D^2+...\,.
\label{Rato}
\end{equation}
The analogous relation for the 1PI function
$\Gamma_{\varphi'\varphi'}$ is
\begin{equation}
\frac{\text{Tr}\,\Gamma_{\varphi'\varphi'}\left|_{ \omega=0}\right.} {g\nu^3\mu^{2\varepsilon}
p^{4-d-2\varepsilon}(d-1)}={Z}_D+
u s^{2\varepsilon}\,
\gamma_{\varphi'\varphi'}^{(1)}\,{Z}_\nu^{-3}{Z}_D^2+(u
s^{2\varepsilon})^2\,\gamma_{\varphi'\varphi'}^{(2)}\,{Z}_\nu^{-6}{Z}_D^3+...\,
. \label{Rat}
\end{equation}
The expansion parameter is
$u=g\bar S_d$ from Eq. (\ref{surface}).
In Sec. \ref{sec:d>2} the quantity
$d$ was considered a fixed parameter and therefore
it was possible to treat $\bar S_d$ as a simple normalization factor.
Here, $d$ is determined by the relation
(\ref{dz}) and in calculations within the usual MS scheme the quantity
$\bar S_d$ should be expanded in the small parameter
$\Delta\sim\varepsilon$. Following Ref. \cite{Ronis},
we shall use the modified  scheme
${\rm \overline{MS}}$ (see, e.g., Ref. \cite{Coll}), in which the
quantity $\bar S_d$ is treated as a whole and not expanded in
$\Delta$. It is well known that the choice of scheme is not reflected
in any physically significant results.

The constants
${Z}$ are sought as series of form
(\ref{1.30}) and determined from the condition of cancellation of the UV divergences
(poles in $\varepsilon$ with
$\Delta/\varepsilon=const$) in relations (\ref{Rato}) and (\ref{Rat}).
Denoting by
${Z}^{(n)}$ the contribution of order $u^n\sim g^n$
in any of these constants we arrive at expressions similar to
(\ref{1z}) and (\ref{2z}):
at the first order in $u\sim g$
\begin{align}
{Z}_\nu^{(1)}&= \mathcal{L}_\eps\,
\left\{us^{2\varepsilon}\gamma_{\varphi'\varphi}^{(1)}\right\}\,,\nonumber\\
{Z}^{(1)}_D&= -\mathcal{L}_\eps\,
\left\{us^{2\varepsilon}\gamma_{\varphi'\varphi'}^{(1)}\right\}\,,
\label{1zz}
\end{align}
and at the second order
\begin{align}
{Z}_\nu^{(2)}&= \mathcal{L}_\eps\,
\left\{u^2s^{4\varepsilon}\gamma_{\varphi'\varphi}^{(2)}+
us^{2\varepsilon}\gamma_{\varphi'\varphi}^{(1)}\,[{Z}^{(1)}_D-
2{Z}_\nu^{(1)}]\right\},\label{2zz}\\
{Z}^{(2)}_D&= \mathcal{L}_\eps\,
\left\{-u^2s^{4\varepsilon}\gamma_{\varphi'\varphi'}^{(2)}
%\right.\nonumber\\&\left.
+us^{2\varepsilon}\gamma_{\varphi'\varphi'}^{(1)}\,[3{Z}_\nu^{(1)}-2{Z}^{(1)}_D]\right\}.
\label{2zzz}
\end{align}
For calculation in the two-loop approximation the following contributions are needed
\begin{align}
\gamma_{\varphi'\varphi}^{(1)}&=\frac{A}{\varepsilon}+B+...,&\gamma_{\varphi'\varphi'}^{(1)}
&=\frac{A'}{\varepsilon}+B'+...,
 \label{ABCD1}\\
\gamma_{\varphi'\varphi}^{(2)}&=\frac{C}{\varepsilon^2}+\frac{D}{\varepsilon}+...,
&\gamma_{\varphi'\varphi'}^{(2)}&=\frac{C'}{\varepsilon^2}+\frac{D'}{\varepsilon}+...\,
. \label{ABCD2}
\end{align}
These are analogs of relations
(\ref{AB}) and (\ref{CD}) with different coefficients, however,
which now may depend on the ratio $\Delta/\varepsilon=\zeta$.

Substituting expressions (\ref{ABCD1}) in Eq. (\ref{1zz}) we find the one-loop contributions to the
renormalization constants:
\begin{eqnarray}
{Z}^{(1)}_\nu=\frac{uA}{\varepsilon}\,,\qquad
{Z}^{(1)}_D=-\frac{uA'}{\varepsilon}\,.
\label{100}
\end{eqnarray}
One-loop calculation yields the following values
(first obtained in Ref. \cite{Ronis}):
\begin{eqnarray}
A=-1\,, \qquad A'=\frac{1}{2+\zeta}\,.
\label{01}
\end{eqnarray}
In the one-loop approximation there are no problems with
$\log s$ in the constants
$Z$, so that the multiplicative renormalization
(\ref{19}) appears quite acceptable.

Consider now two-loop contributions
(\ref{2zz}) and (\ref{2zzz}). Taking into account the already
known one-loop expressions
(\ref{100}) we obtain
\begin{align}
{Z}_\nu^{(2)}&=\mathcal{L}_\eps\,
\left\{u^2s^{4\varepsilon}\left(\frac{C}{\varepsilon^2}+\frac{D}{\varepsilon}\right)
%\right.\nonumber\\ &+\left.
+
us^{2\varepsilon}\left(\frac{A}{\varepsilon}+B\right)\left(-\frac{uA'}{\varepsilon}-\frac{2uA}{\varepsilon}\right)
\right\},\label{3zz}\\
{Z}^{(2)}_D&=\mathcal{L}_\eps\,
\left\{-u^2s^{4\varepsilon}\left(\frac{C'}{\varepsilon^2}+\frac{D'}{\varepsilon}\right)
%\right.\nonumber\\&+\left.
+
us^{2\varepsilon}\left(\frac{A'}{\varepsilon}+B'\right)\left(\frac{3uA}{\varepsilon}+\frac{2uA'}
{\varepsilon}\right)\right\}.
\label{3zzz}
\end{align}
The condition of cancellation of the contributions
$\sim\varepsilon^{-1} \log s$ in Eq. (\ref{3zz}) is
\begin{eqnarray}
4C+2A(-A'-2A)=0,
\label{usl1}
\end{eqnarray}
and in Eq. (\ref{3zzz}) analogously
\begin{eqnarray}
-4C'+2A'(3A+2A')=0.
\label{usl2}
\end{eqnarray}
Our two-loop calculation of the coefficients
$C$ and $C'$ yields
\begin{align}
C&=1-\frac{1}{2(2+\zeta)},\nonumber\\
C'&=\frac{2}{(2+\zeta)(3+\zeta)}-\frac{3}{(3+\zeta)}.
\label{001}
\end{align}
Substitution in relations
(\ref{usl1}) and (\ref{usl2}) of the calculated quantities
(\ref{01}) and (\ref{001}) readily shows that condition
(\ref{usl1}) is satisfied, whereas  (\ref{usl2}) is not. This means that
in ${Z}_\nu^{(2)}$ there is no ''bad'' contribution
$\sim\varepsilon^{-1}\log s = \varepsilon^{-1}\log(\mu/p)$, while
in $ {Z}^{(2)}_D$ there is such a term:
\begin{eqnarray}
\frac{2(1+\zeta)(4+3\zeta)}{(2+\zeta)^2(3+\zeta)}\cdot
\varepsilon^{-1} \log (\mu/p),
\label{002}
\end{eqnarray}
whose coefficient is the expression of the left-hand-side of Eq. (\ref{usl2}).

Thus within the renormalization scheme of Ref. \cite{Ronis} according to relations (\ref{19})
a dependence on the external wave numbers through
$\log s=\log (\mu/p)$ appears in the renormalization constants, which is completely inacceptable by the very
definition of the renormalization constants. It is not difficult to understand the reason of this: in
scheme (\ref{19}) there is a violation of a fundamental principle of the general theory of UV renormalization --
requirement that all counterterms must be local (polynomial functions of external wave vectors) \cite{Coll}.
The introduction of the coefficient
${Z}_D$ at the term
$\sim\varphi'D\varphi'$ in action (\ref{action})
with the nonlocal injection function (\ref{nakach}) is tantamount to introduction of
a nonlocal counterterm with the structure $p^{4-d-2\varepsilon }$. This feature takes
the scheme discussed beyond the framework of the standard theory of UV renormalization
with such unpleasant consequences as the appearance of the (inacceptable) dependence on wave numbers in the
renormalization constants. This general line of argument motivated the authors of Ref. \cite{Nalimov}
to change the scheme of
($\varepsilon$, $\Delta$) renormalization to conform to the requirement of the polynomial in wave vectors
form of all the counterterms (localness), although in the one-loop calculation of Ref. \cite{Ronis} the
inconsistency of the scheme proposed there does not show explicitly.

It might be suggested to change relation
(\ref{2zzz}) to exclude the wave-number-dependent contribution (\ref{002}) from
${Z}_D^{(2)}$. Eq. (\ref{2zzz}) was obtained, however, from the requirement that
in the two-loop approximation all UV divergences -- poles in $\eps$ -- were removed from the renormalized
1PI function
$\Gamma_{\varphi'\varphi'}$, so that any change of the form of
${Z}_D^{(2)}$  from (\ref{2zzz})
would lead to the appearance of poles in $\varepsilon$
in the renormalized function
$\Gamma_{\varphi'\varphi'}$.

The persistent opponent might say:''Who cares, I am not interested in the two-loop approximation, I am
completely happy with the one-loop accuracy, where there are no problems.'' Here, the objection
would be that elimination of UV divergences (poles in
$\varepsilon$) to all orders in perturbation theory is not a caprice but a compelling necessity.
If such poles are left, then there is no guarantee
that results obtained at the lowest order of perturbation theory do not acquire corrections of the same order
from the higher-order terms not accounted for (in fact, there is conviction in the opposite),
i.e. lowest-order calculations become completely unreliable.
Therefore, in particular, the conclusion of Ref. \cite{Ronis} that relations
(\ref{razm}) are violated in the
($\varepsilon$, $\Delta$) scheme is not correct; in the consistent renormalization scheme these relations
continue to hold \cite{Nalimov}.

Let us briefly discuss the possibility to carry out a nonlocal
renormalization in the NP scheme. Relations (\ref{19}) are
preserved in this case, whereas the two independent renormalization constants
$ Z_\nu$ and $ Z_{D_2}$ are to be determined
from the following normalization conditions at $p=\mu$ for the 1PI
functions $\Gamma_{\varphi'\varphi}$ (\ref{Rato}) and
$\Gamma_{\varphi'\varphi'}$ (\ref{Rat}):
\begin{align}
\frac{\text{Tr}\,\Gamma_{\varphi'\varphi}|_{ \omega=0}} {\nu
p^2(d-1)}\biggl|_{p=\mu} &=-1 \,, \nonumber\\
\frac{\text{Tr}\,\Gamma_{\varphi'\varphi'}\left|_{ \omega=0}\right.}
{g\nu^3\mu^{2\varepsilon} p^{4-d-2\varepsilon}(d-1)}
\biggl|_{p=\mu} &=1. \label{Rato1}
\end{align}
The problem of dependence of the renormalization constants on
wave number is absent in such a setup. However, it may be readily
checked that conditions (\ref{usl1}) and (\ref{usl2}) remain
necessary to ensure absence of UV-divergent (at $\eps\to 0$) contributions $\sim
u^2\varepsilon^{-1} \log (\mu/p)$ in the renormalized Green
functions $\Gamma_{\varphi'\varphi}$ and
$\Gamma_{\varphi'\varphi'}$ for arbitrary values of the wave number $p$.

In conclusion, let us point out that the ''bad'' contribution (\ref{002}) in
${Z}_D^{(2)}$ vanishes at $\zeta=-1$, i.e.
at $\Delta=-\varepsilon$ in Eq. (\ref{dz}). Then $d=2+2\Delta=2-2\varepsilon$
and energy injection (\ref{nakach}) becomes local: $d_f\sim p^{4-d-2\varepsilon}=p^2$
(such a model was considered in Ref. \cite{Nelson}). In this case the multiplicative renormalization
(\ref{19}) conforms to the requirement of local counterterms and the corresponding
constants ${Z}$ do not contain any dependence on
$\log s$ in accordance with the general theory.

\section{\label{sec:local}Construction of the ($\eps$, $\Delta$) expansion in the two-charge model with
local counterterms. Two-loop calculation of the renormalization constants}

In the preceding section it was shown that in the
($\eps$ ,$\Delta$) scheme
(\ref{dz}) the multiplicative renormalization \cite{Ronis}
of the amplitude $D_0$ in Eq. (\ref{nakach}) is not acceptable. The reason
is that the counterterm with structure (\ref{nakach})
is nonlocal $\sim k^{4-d-2\eps}=k^{2-2\Delta-2\eps}$ on rays (\ref{dz}).

Guided by the general theory of the UV renormalization, the authors of Ref. \cite{Nalimov}
put forward another scheme, in which a local counterterm  $\sim k^2$ instead of the nonlocal
one $ \sim k^{2-2\Delta-2\eps}$ is used to absorb singularities from the graphs of
the 1PI function $\Gamma_{\varphi'\varphi'}$.
This corresponds to addition of the term
$\sim \varphi'\partial^2\varphi'$ to the action functional. In functional (\ref{action})
with the correlation function $D$ from Eqs. (\ref{1.2}) and (\ref{nakach}) there is no such term,
so that upon the addition of the term
$\sim \varphi'\partial^2\varphi'$ the renormalization ceases to be multiplicative.
This would be unessential, if our only goal was the elimination of divergences from
Green's functions which is quite possible by a non-multiplicative renormalization. For the
use of the standard technique of the RG multiplicative renormalization is, however, necessary.
This is why the authors of Ref. \cite{Nalimov} proposed to consider a two-charge model in which
to function (\ref{nakach})
$\sim k^{4-d-2\eps}=k^{2-2\Delta-2\eps}$ the term
$\sim k^2$ is added at the outset with an independent coefficient:
\begin{equation}
    d_f(k)=D_{10} k^{2-2\Delta-2\eps}+D_{20} k^2
    = g_{10}  \nu_0^{3}\,k^{2-2\Delta-2\eps}+g_{20} \nu_0^{3}\, k^2\,.
\label{nakach2}
\end{equation}
Here, the amplitude
$D_0$ of Eq. (\ref{nakach}) is denoted by $D_{10}$. The parameters  $g_{10}$ and $g_{20}$
introduced in Eq. (\ref{nakach2}) play the role of two independent bare charges.

The contribution with  $D_{20}$ in relation (\ref{nakach2}) corresponds to thermal fluctuations.
A model with this term only has been analyzed earlier in Ref. \cite{Nelson}. In the theory
of turbulence $D_{20}=0$ should be considered the ''real value'' of this parameter, since only the
first term in Eq. (\ref{nakach2}) at $\eps=2$ reproduces the realistic for the theory of turbulence
pumping of energy by large-scale eddies. It will be shown below that vanishing of the bare
parameter $g_{20}=D_{20} \nu_0 ^{-3} = 0$ does not imply vanishing of the corresponding renormalized
parameter $g_2$, so that in terms of renormalized parameters function (\ref{nakach2}) gives rise to
a two-charge model.

The unrenormalized action is, as before, functional (\ref{action}), but now with the injection function
(\ref{nakach2}) instead of (\ref{nakach}) in correlation function (\ref{1.2}). In the adopted shorthand
notation
\begin{equation}
%\nonumber
 S(\Phi )={1\over 2}\varphi '(D_{1 0}k^{2-2\Delta-2\eps}+D_{2 0}k^{2} )\varphi '
 +\varphi '[-\partial
_t\varphi +\nu _0\partial^{2} \varphi -(\varphi
\partial )\varphi ] \,. \label{action3}
\end{equation}
The propagators $\langle \varphi\varphi'\rangle _0$ and
$\langle \varphi'\varphi'\rangle _0 $ corresponding to action
(\ref{action3}) maintain earlier form (\ref{lines}), whereas $\langle
\varphi\varphi\rangle _0$ is replaced by
\begin{equation}
\langle \varphi \varphi \rangle _0=\frac{(D_{1
0}k^{2-2\Delta-2\eps}+D_{2 0}k^{2} )}{ 2\nu_0 k^{2}}
\exp
\left[-\nu_0 k^{2} |t-t'| \right]\,. \label{lines2}
\end{equation}
We are interested in the region $\eps>0$ and $\Delta>0$
in Eq. (\ref{dz}). In this region in model (\ref{action3}) the additional problem of ''$\Lambda$ divergences''
arises which was absent in model (\ref{action}) with injection function (\ref{nakach}).
Let us explain this in more detail. Wave-vector integrals --
with the shorthand notation $\int dk\dots$ -- corresponding to the 1PI
graphs discussed always reduce to
''nearly logarithmic'' ones in the present set of models. Their deviation from
logarithmicity shows in the form of factors of the type
$k^{\alpha}$ with a small exponent $\alpha=2m \Delta-2n \eps$, where $n$ and $m$
are nonnegative integers.
The exponent $\alpha$ is the wave-number dimension
of the wave-vector integrals obtained upon all time integrations and may be calculated
by the following simple rule: each loop integral over wave vectors contributes a term
$2\Delta$ to $\alpha$, the term with $D_{10}$ in Eq. (\ref{lines2}) yields the contribution
$-2\eps-2\Delta$, but the term with $D_{20}$ does not affect $\alpha$ at all.
Thus it may readily be seen that if only the nonlocal term with $D_{10}$ is left in
Eq. (\ref{nakach2})
[i.e. if we return to model (\ref{nakach})], then all the exponents $\alpha$
in graphs of $\Gamma_{\varphi'\varphi}$ and $\Gamma_{\varphi'\varphi'}$
at $\eps>0$ and  $\Delta >0$ become negative. All the integrals in the limit
$k\to\infty$ converge, they may be carried out over the whole wave-vector space and
the divergences show as poles in $\eps$, $\Delta$ and their linear combinations.

However, in the model with injection
(\ref{nakach2}) -- due to the presence of the second term with $D_{20}$  --
at $\Delta>0$ wave-vector integrals appear with $\alpha >0$. They diverge in the limit
$k\to\infty$ and thus require an UV cutoff $\Lambda$. As examples we quote
the values of $\alpha$ in the graphs of interest for us.
In the one-loop graphs of $ \Gamma_{\varphi'\varphi}$: $ \alpha=-2\eps$, $2\Delta$;
in the two-loop graphs: $\alpha= -4\eps$, $-2\eps+2\Delta$, $4\Delta$; in the one-loop graphs
of $\Gamma_{\varphi'\varphi'}$: $\alpha= -4\eps-2\Delta$, $-2\eps$, $2\Delta$; in the
two-loop graphs: $\alpha=-6\eps-2\Delta$, $-4\eps$, $-2\eps+2\Delta$, $4\Delta$.

Thus, in the two-charge model (\ref{nakach2}) at $\Delta>0$ ($\eps>0$ is always implied)
some integrals have the $\Lambda$ divergence at large $k$. To remove these divergences
an additional procedure of $\Lambda$ renormalization procedure is needed which we shall discuss
in the Appendix. At the moment the important point is that after the $\Lambda$ renormalization
the limit $\Lambda\to\infty$ may be taken with the result that divergences appear only in the
form of poles in $\eps$, $\Delta$ and their linear combinations. The same poles may be found
within the ''formal scheme'', where all integrals are understood as analytic continuation on the
parameters $\eps$ and $\Delta$ from the region, where there are no $\Lambda$ divergences.

In our case this is the region of $\eps>0$ and small (compared with $\eps$) negative $\Delta<0$
(i.e. $d<2$).
In this section we shall consider results obtained in the framework of this ''formal scheme''.
There is no UV-cutoff parameter $\Lambda$ in this scheme, but the divergences appear in the form
of poles in $\eps$ with $\Delta/\eps=const$. The goal of the renormalization is
removal of these poles. In the Appendix it will be shown that the results obtained this way
coincide with those obtained at $\Delta>0$ after the $\Lambda$ renormalization and subsequent
limit $\Lambda\to\infty$.

The relations of multiplicative renormalization in the formal scheme are
\begin{alignat}{2}
D_{10}& = g_{10}\nu_0^{3} =
g_1\mu^{ 2\eps} \nu^{3},
&g_{10}&=g_1\mu^{2\eps}{Z}_{g_1},
\nonumber\\
D_{20} &= g_{20}\nu_0^{3} =
g_2\mu^{-2\Delta} \nu^{3}{Z}_{D_2},\!\!\!
&g_{20}&=g_2\mu^{-2\Delta}{Z}_{g_2},
\label{ZZ}\\
\nu_0&=\nu {Z}_{\nu},\quad\
{Z}_{g_1}{Z}_{\nu}^{3}=1, &{ Z}_{g_2}{Z}_{\nu}^{3}&={Z}_{D_2}\,,\nonumber
\end{alignat}
with two independent renormalization constants for the viscosity
${\nu_0}$ and the amplitude ${D_{20}}$; the amplitude ${D_{10}}$ of the nonlocal
correlation function of the random force is not renormalized. The renormalization
constants $Z_{\nu}$ and $Z_{D_2}$ are found from the condition that the 1PI functions
$\Gamma_{\varphi'\varphi}\left|_
{\omega=0}\right.$ and $\Gamma_{\varphi'\varphi'}\left|_ {\omega=0}\right.$ are UV finite
(i.e. with $\Delta/\eps=const$ there are no poles in $\eps$).
The dimensionless expansion parameters of the perturbation theory for these quantities are
\begin{equation}
\alpha_1\equiv\frac{D_{10}\overline{S}_d}{32\nu_0^3p^{2\eps}}\,
,\qquad
\alpha_2\equiv\frac{D_{20}\overline{S}_d}{32\nu_0^3p^{-2\Delta}}
\label{alpha}
\end{equation}
with $\overline{S}_d$ from Eq. (\ref{surface}).
Instead of relation (\ref{ti}) we now have
\begin{equation}
{\text{Tr}\,\Gamma_{\varphi'\varphi}\left|_
{\omega=0}\right. \over d-1}= \nu_0
p^2\left[-1+\negthickspace\negthickspace\negthickspace
\sum\limits_{n_1\geq 0,n_2\geq0,\atop n_1+n_2\geq 1}\!\negthickspace\negthickspace
\alpha_1^{n_1}\alpha_2^{n_2}
\gamma^{(n_1,n_2)}_{\varphi'\varphi}\right]\,,
\label{ti3}
\end{equation}
and the analogous expression for $\Gamma_{\varphi'\varphi'}$:
\begin{equation}
{\text{Tr}\,\Gamma_{\varphi'\varphi'}\left|_ {\omega=0}\right.\over d-1} = D_{10} p^{2-2\Delta-2\eps}
+D_{20}p^2
\left[1+\negthickspace\negthickspace\negthickspace\negthickspace
\sum\limits_{n_1\geq0,n_2\geq-1,\atop n_1+n_2\geq1}
\!\negthickspace\negthickspace
\alpha_1^{n_1}\alpha_2^{n_2}
\gamma^{(n_1,n_2)}_{\varphi'\varphi}\right]\,. \label{ti4}
\end{equation}
In terms of the renormalized variables relations (\ref{ti3}) and (\ref{ti4})
yield for the reduced dimensionless functions the following representations
\begin{equation}
{\text{Tr}\,\Gamma_{\varphi'\varphi}\left|_{ \omega=0}\right.\over \nu p^2(d-1)}
=-{Z}_\nu+
{Z}_\nu\negthickspace\negthickspace\negthickspace\negthickspace\negthickspace
%\negthickspace\negthickspace\negthickspace\negthickspace
\sum\limits_{n_1\geq0,n_2\geq 0,\atop n_1+n_2\geq1}
\!\negthickspace\negthickspace
\alpha_1^{n_1}\alpha_2^{n_2}
\gamma^{(n_1,n_2)}_{\varphi'\varphi}, \label{Rato2}
\end{equation}

\begin{equation}
{\text{Tr}\,\Gamma_{\varphi'\varphi'}\left|_{ \omega=0}\right.\over (d-1)g_2\nu^3\mu^{-2\Delta}
p^{2}}
=\frac{u_1}{u_2}s^{2\eps+2\Delta}+Z_{D_2}
+Z_{D_2}\negthickspace\negthickspace\negthickspace\negthickspace
\sum\limits_{n_1\geq0,n_2\geq -1,\atop n_1+n_2\geq1}
\!\negthickspace\negthickspace
\alpha_1^{n_1}\alpha_2^{n_2}
\gamma^{(n_1,n_2)}_{\varphi'\varphi'}\, , \label{Rat2}
\end{equation}
where the expansion parameters $\alpha_1$ and $\alpha_2$ from Eq. (\ref{alpha}) are
expressed through the renormalized parameters according to relations (\ref{ZZ}):
\begin{equation}
\alpha_1 =
{u_1 s^{2\eps}}{Z_{\nu}^{-3}},\qquad
\alpha_2 =
{u_2 s^{-2\Delta}Z_{D_2}}{Z_{\nu}^{-3}}
\label{alpha2}
\end{equation}
Here, $u_1=g_1 \overline{S}_d/32$, $u_2=g_2 \overline{S}_d/32$ and
$s\equiv\mu/p$.
Dependence on $\varepsilon$ of the coefficient functions $\gamma^{(n_1,n_2)}_{\varphi'\varphi}$ and
$\gamma^{(n_1,n_2)}_{\varphi'\varphi'}$ in Eqs. (\ref{Rato2}) and (\ref{Rat2})
is determined  by relations of the form of Eqs. (\ref{ABCD1}) and (\ref{ABCD2}),
in which the
$\zeta=\Delta/\eps$-dependent coefficients $A$, $B$, $C$, $D$, $A'$, $B'$, $C'$ and $D'$
now acquire subscripts corresponding to the superscripts
($n_1,n_2$) of the quantities $\gamma^{(n_1,n_2)} $. In the one-loop approximation
the following analogs of relations (\ref{ABCD1}) are needed
  \begin{equation}
   \gamma_{\varphi'\varphi}^{(i,k)}=\frac{A_{i,k}}{\eps}+B_{i,k}\,,
    \qquad\gamma_{\varphi'\varphi'}^{(i,k)}=\frac{A'_{i,k}}{\eps}+B'_{i,k}
   \,,
\label{ABCD3}
 \end{equation}
with the index sets
$(i,k)=\{(1,0),(0,1)\}$ for $\gamma_{\varphi'\varphi}$
and the sets $(i,k)=\{(2,-1),(1,0),(0,1)\}$ for $\gamma_{\varphi'\varphi'}$.
In the two-loop approximation the following analogs of relations (\ref{ABCD2})
have to be included
  \begin{equation}
   \gamma_{\varphi'\varphi}^{(i,k)}=\frac{C_{i,k}}{\eps^2}+\frac{D_{i,k}}{\eps}\,,
   \qquad\gamma_{\varphi'\varphi'}^{(i,k)}=\frac{C'_{i,k}}{\eps^2}+\frac{D'_{i,k}}{\eps}
   \,,
\label{ABCD4}
 \end{equation}
with the sets
$(i,k)=\{(2,0),(1,1),(0,2)\}$ for $\gamma_{\varphi'\varphi}$
and $(i,k)=\{(3,-1),(2,0),(1,1),(0,2)\}$ for $\gamma_{\varphi'\varphi'}$.
For calculation of the constants
$Z$ in the one-loop approximation, which was carried out in Ref. \cite{Nalimov},
only constants $A$ and $A'$ from Eq. (\ref{ABCD3}) are needed. For our two-loop
calculation all constants in Eqs. (\ref{ABCD3}) and (\ref{ABCD4}) are necessary.

The constants  $ {Z_\nu}$ and $Z_{D_2}$ are determined from the condition of cancellation of
all UV divergences ( poles in $\varepsilon$ with $\Delta/\varepsilon=const$) in Eqs. (\ref{Rato2}) and
(\ref{Rat2}). Denoting by ${Z}^{(n)}$ the contribution of order $u^n\sim g^n$
with respect to the set of charges $u_1$ and $u_2$ to any constant we obtain
at the first order in $u\sim g$
\begin{align}
\nonumber {Z}_\nu^{(1)}&= \mathcal{L}_\eps\,
\left[u_1s^{2\varepsilon}\gamma_{\varphi'\varphi}^{(1,0)}+u_2s^{-2\Delta}\gamma_{\varphi'\varphi}^{(0,1)}\right],\\
{Z}^{(1)}_{D_2}&= -\mathcal{L}_\eps\,
\left[\frac{u_1^2}{u_2}s^{4\varepsilon+2\Delta}\gamma_{\varphi'\varphi'}^{(2,-1)}
%\right.\\&\left.
+{u_1}s^{2\varepsilon}\gamma_{\varphi'\varphi'}^{(1,0)}+{u_2}s^{-2\Delta}\gamma_{\varphi'\varphi'}^{(0,1)}\right],
\label{1zz2}
%\nonumber
\end{align}
and at the second order
\begin{widetext}
\begin{multline}
{Z}_\nu^{(2)}=
\mathcal{L}_\eps\,\left\{u_1^2s^{4\varepsilon}\gamma_{\varphi'\varphi}^{(2,0)}+
u_1 u_2s^{2\varepsilon-2\Delta}\gamma_{\varphi'\varphi}^{(1,1)}+
u_2^2s^{-4\Delta}\gamma_{\varphi'\varphi}^{(0,2)}
\right.\\
\left.
+u_1s^{2\eps}\gamma_{\varphi'\varphi}^{(1,0)}\,\left[-2{Z}_\nu^{(1)}\right]+
u_2s^{-2\Delta}\gamma_{\varphi'\varphi}^{(0,1)}\,\left[{Z}^{(1)}_{D_2}-
2{Z}_\nu^{(1)}\right]\right\},\label{2zz2}
\end{multline}
\begin{multline}
 {Z}^{(2)}_{D_2}=-
\mathcal{L}_\eps\,\left\{\frac{u_1^3}{u_2}s^{6\varepsilon+2\Delta}\gamma_{\varphi'\varphi'}^{(3,-1)}+
{u_1^2}s^{4\varepsilon}\gamma_{\varphi'\varphi'}^{(2,0)}+
{u_1}{u_2}s^{2\varepsilon-2\Delta}\gamma_{\varphi'\varphi'}^{(1,1)}+
{u_2^2}s^{-4\Delta}\gamma_{\varphi'\varphi'}^{(0,2)}\right.\\
\label{2zzz2}\left.+
\frac{u_1^2}{u_2}s^{4\varepsilon+2\Delta}\gamma_{\varphi'\varphi'}^{(2,-1)}\,\left[-3{Z}_\nu^{(1)}\right]+
u_1s^{2\varepsilon}\gamma_{\varphi'\varphi'}^{(1,0)}\,\left[{Z}^{(1)}_{D_2}-3{Z}_\nu^{(1)}\right]
+
u_2s^{-2\Delta}\gamma_{\varphi'\varphi'}^{(0,1)}\,\left[2{Z}^{(1)}_{D_2}-3{Z}_\nu^{(1)}\right]\right\}.
\end{multline}
Substituting expressions
(\ref{ABCD3}) and (\ref{1zz2}) we find the one-loop contributions to the renormalization
constants
\end{widetext}
\begin{equation}
%\nonumber
{Z}_\nu^{(1)}=\frac{1}{\eps} \left( u_1 A_{1,0}+u_2
A_{0,1}\right),
{Z}^{(1)}_{D_2}
= -\frac{1}{\eps}\left(
\frac{u_1^2}{u_2}A'_{2,-1}+u_1A'_{1,0}+{u_2}A'_{0,1}\right),
\label{1zz22}
\end{equation}
The coefficients $A$ and $A'$ here have been calculated in Ref. \cite{Nalimov}. In our notation
\begin{align}
 A_{1,0}&=-1\,, &A_{0,1}&=\frac{1}{\zeta}\,,&A'_{2,-1}&=\frac{1}{2+\zeta}\,,\nonumber\\
 A'_{1,0}&=2\,,
 &A'_{0,1}&=-\frac{1}{\zeta}\,.&&
 \label{Aform}
\end{align}
In the present work we have carried out the two-loop calculation and determined the
coefficients $B$ and $B'$ in Eq. (\ref{ABCD3}) together with $C$, $C'$, $D$ and $D'$
in Eq. (\ref{ABCD4}).
Let us quote the coefficients  $C$ and $C'$ necessary at the moment:
\begin{align}
C_{2,0}&=1-\frac{1}{2(2+\zeta)}\,,\
C_{1,1}=-\frac{2}{\zeta(1-\zeta)}\,,\
C_{0,2}=\frac{1}{2\zeta^2}\,,\nonumber\\
C'_{3,-1}&=\frac{2}{(2+\zeta)(3+\zeta)}-\frac{3}{3+\zeta}\,,\
C'_{0,2}=-\frac{1}{2\zeta^2}\,,\label{Cform}\\
C'_{2,0}&=-1+\frac{1}{2+\zeta}+\frac{3}{2\zeta}\,,\quad
C'_{1,1}=\frac{4}{\zeta(1-\zeta)}+\frac{1}{1-\zeta}\,.
\nonumber
\end{align}
To check the cancellation of the ''bad'' terms  $\sim \eps^{-1} \log s$ in Eqs. (\ref{2zz2})
and (\ref{2zzz2}) only terms $\sim 1/\eps^2$ from them are needed. They are determined in Eq. (\ref{ABCD4}) by
the coefficients $C$ and $C'$ from Eq. (\ref{Cform}) and in
the contributions with $A$ and $A'$ from Eqs. (\ref{ABCD3}),
(\ref{1zz22}) and (\ref{Aform}). Substitution shows that all contributions
with $\eps^{-1}\log s$ in Eqs. (\ref{2zz2}) and (\ref{2zzz2}) cancel as required.

A specific feature of the renormalization constant $Z_{D_2}$
is that it contains terms $\sim1/u_2$ [see Eq. (\ref{1zz22})]. When such a $Z_{D_2}$
is substituted in renormalization relations (\ref{ZZ}) in the expression for $D_{20}$ terms
independent of $u_2$ appear (generation terms):
\begin{equation}
D_{20}\frac{ \overline{S}_d}{32 }=u_2 \mu^{-2\Delta}\nu^3
Z_{D_2}=\nu^3\mu^{-2\Delta}\Big[ u_2
- \frac{1}{\eps}\big(u_1^2
A'_{2,-1}+u_1 u_2 A'_{1,0}+u_2^2A'_{0,1}\big)+\dots\Big]\;.
\label{gener}
\end{equation}
Due to such terms the condition $D_{20}=0$ does not lead to the trivial
conclusion $u_2=0$, i.e. a nonvanishing value of the renormalized charge
corresponds even to the zero (real) value of the bare charge.

The ellipsis in Eq. (\ref{gener}) stands for contributions of the two-loop order and higher, which
contain terms $\sim u^n/\varepsilon^{n-1}$ with $n\geq 3$. In the region $u \sim \varepsilon$
(where the fixed point $u_*$ of the RG lies, see Sec. \ref{sec:RGE}) they are of the same order
in $\eps$ as the explicitly quoted one-loop contribution in Eq. (\ref{gener}). Therefore, to determine
the connection between the charges $u_1$ and $u_2$ imposed by the condition $D_{20}=0$ (i.e.
$Z_{D_{2}}=0$) the two-loop calculation of the constants $Z$ is not sufficient. This is unimportant,
however, in the following, because in the RG analysis of Sec. \ref{sec:RGE} the charges
$u_1$ and $u_2$ are considered independent parameters.

We shall not quote here the fairly cumbersome expressions obtained
by us for the constants $B$, $B'$, $D$ and $D'$ in Eqs. (\ref{ABCD3}) and
(\ref{ABCD4}). Instead, we quote the two-loop expressions for the
renormalization constants $Z_{\nu}$ and $Z_{D_2}$ obtained with the use of
them and relations (\ref{2zz2}) - (\ref{Cform}) in the ${\rm \overline{MS}}$
scheme (a detailed account of the method of calculation can be found in Ref. \cite{Adzhemyan03a}):
\begin{widetext}
\begin{multline}\label{Znu}
Z_{\nu}=1 -{\frac { u_{{1}}}{\epsilon\, }}
+{\frac{u_{{2}}}{\zeta\,\epsilon}} -\frac{1}{2}\left[\,{\frac
{4\,\zeta+3}{ \left( 2+\zeta \right)
\epsilon}}+\,{\frac{2\,\zeta+1}{\zeta\,{\epsilon}^{2}}} \right]
{u_{{1}}}^{2}
 - \left[ \,{\frac {5\,\zeta+3 }{\epsilon\, \left( 1-\zeta \right)}}+\,{\frac {2}
 { \left(1- \zeta\right) {\epsilon}^{2}}} \right]
u_{{1}}u_{{2}}\\ -\frac{1}{2} \left[ \,{\frac
{1}{\zeta\,\epsilon}}+\,{\frac {1}{{\zeta}^{2} {\epsilon}^{2}}}
\right] {u_{{2}}}^{2} + \frac{1}{{\epsilon}}\,\left[
{{{u_{{1}}}^{2}}} +4{\frac{u_{{1}}u_{{2}}}{ \left( 1-\zeta \right)
}} -{\frac{{u_{{2}}}^{2}}{\zeta} } \right] R,
\end{multline}
\begin{multline}
\label{ZD}
Z_{D_2}= 1 -{\frac {{u_{{1}}}^{2}}{\epsilon\,u_{{2}}
\left( 2+\zeta
 \right) }}
-\frac {2 u_{{1}}}{\epsilon}
 +{\frac {u_{{2}}}{\zeta\,\epsilon}}
 + \left[ \,{\frac {\zeta\, \left( 13+19\,\zeta \right) }
{2 \left ( 3+\zeta \right)  \left( 2+\zeta \right) \epsilon}} +{
\frac {2\,\zeta+1}{ \left( 3+\zeta \right) \left( 2+\zeta \right)
{ \epsilon}^{2}}} \right] {u_{{1}}}^{3}{u_{{2}}}^{-1}\\
 -\frac{1}{2}\left[ \,{\frac {34\,\zeta+19+6\,
{\zeta}^{2}}{ \left( 2+\zeta \right) \epsilon}}+
\,{
\frac { \left( \zeta+4 \right)  \left( 2\,\zeta+1 \right)
}{\zeta\,
 \left( 2+\zeta \right) {\epsilon}^{2}}} \right]
{u_{{1}}}^{2} - \frac{1}{2}\left[ \,{\frac {13+31\,
\zeta}{\epsilon\, \left( 1-\zeta \right) }}+\,{\frac
{2(4\,\zeta+1)}{
 \left( 1-\zeta \right) \zeta\,{\epsilon}^{2}}} \right]
 u_{{1}}u_{{2}}
  -\frac{1}{2}\left( {\frac {3}{\zeta\,\epsilon}}+{\frac {1}{{\zeta}^{2}
{\epsilon}^{2}}} \right) {u_{{2}}}^{2}\\  +
\frac{1}{\,\epsilon}\,\left[ {2}{\frac {{u_{{1}}}^{3}}{u_{{2}}
\left( 3+\,\zeta
 \right) }}+{3}\,{{{u_{{1}}}^{2}}}+6\,{\frac {u_{{1}}u_{{2}}}{
 \left( 1-\,\zeta \right) }} -{\frac {{u_{{2}}}^{2}}{\zeta}}\right] {(\it R-1)}\,,
\end{multline}
\end{widetext}
where
$$R=-0.168\,.$$
This number has been obtained by a computer calculation of a relatively simple but cumbersome
twofold integral, through which all the nontrivial two-loop contributions in Eqs. (\ref{Znu}) and (\ref{ZD})
are expressed.

In conclusion, we quote the analogs of relations (\ref{Znu}) and (\ref{ZD}) in the NP scheme.
In this scheme the constants $Z$ are determined instead of Eq. (\ref{Rato1}) by normalization conditions
\begin{align}
\label{Rato3}
\frac{\text{Tr}\,\Gamma_{\varphi'\varphi}|_{ \omega=0}} {\nu p^2(d-1)}
\Bigg|_{p=\mu} &=-1\,,\nonumber\\
\frac{\text{Tr}\,\Gamma_{\varphi'\varphi'}\left|_{ \omega=0}\right.}
{g\nu^3\mu^{2\varepsilon} p^{4-d-2\varepsilon}(d-1)}
\Bigg|_{p=\mu}&=\frac{u_1}{u_2}+1\,,
\end{align}
with $\Gamma_{\varphi'\varphi}$ from  Eq. (\ref{Rato2}) and
$\Gamma_{\varphi'\varphi'}$ from Eq. (\ref{Rat2}).
From here the renormalization constants follow in the form:
\begin{widetext}
\begin{multline}\label{Znu1}
 Z_{\nu}=1+ \left( -\frac{1}{\epsilon}-\frac{3}{2}\,\zeta+c \right)
u_{{1}}+ \left( {\frac {1}{\zeta\,\epsilon}}+c+\frac{3}{ 2} \right)
u_{{2}} \\ +\left[\left( -1-\frac{1}{2{\zeta}}
\right)\frac{1}{\epsilon^{2}}+ \left( 2\,c-3\,\zeta+R-2+ \,{\frac
{c-1}{\zeta}} \right) \frac{1}{\epsilon} \right] {u_{{1}}}^{2 } \\
 + \left[ \,{\frac {2}{\left( \zeta-1 \right)
{\epsilon}^{2}}}+
 \left( 6+2\,c-\frac{2}{\zeta}-4\,{\frac {\,R-2}{\zeta-1}} \right)\frac{1} {
\epsilon} \right] u_{{1}}u_{{2}}+ \left( \,{\frac {1}{2{\zeta}
^{2}{\epsilon}^{2}}}-{\frac {2+R+c}{\zeta\,\epsilon}} \right)
{u_{{2}} }^{2}\,,
\end{multline}
\begin{multline}\label{ZD1}
 Z_{D_2}=1+ \left[c -{\frac {1}{ ( 2+\zeta )
\epsilon}}-\frac{7}{2}+\frac{5}{ ( 2+\zeta )} \right]
\frac{u_{{1}}^2}{u_{{2}}}
 + \left(2\,c -\frac {2} {\epsilon}-5\,\zeta-2 \right) u_{{1}}+ \left( c+
\frac{3}{2}+{\frac {1}{\zeta\,\epsilon}} \right) u_{{2}}
\\ + \left[ \left( \frac{5}{ 3+\zeta }-\frac{3}{ 2+\zeta }
\right) \frac{1}{\epsilon^2}+ \left( 12-2\,c+{\frac
{-68+2\,R}{3+\zeta}}+3 \,{\frac {c+8}{2+\zeta}} \right)\frac{1}
{\epsilon} \right] \frac{u_{{1}}^3}{u_{{2}}}\\ + \left[  \left(
2\,c-1-\frac{1}{\zeta}-\frac{3}{ 4+2\zeta} \right)
\frac{1}{\epsilon^2}+ \left( 3\,R-2\,\zeta-10+ {\frac
{2\,c-2}{\zeta}}+3\,{\frac {4+c}{2+\zeta}} \right) \frac{1}{
\epsilon} \right] {u_{{1}}}^{2}\\  + \left[  \left(
\frac{5}{\zeta-1}-\frac{1}{\zeta} \right) \frac{1}{\epsilon^2}+
\left( 16+4\,c+ {\frac {c-4}{\zeta}}+{\frac {28-6\,R}{\zeta-1}}
\right) \frac{1}{\epsilon} \right] u_{{1}}u_{{2}}+ \left( -{\frac
{1}{2{\zeta}^{2}{ \epsilon}^{2}}}-{\frac {2+R+c}{\zeta\,\epsilon}}
\right) {u_{{2}}}^{2}\,,
\end{multline}
\end{widetext}
where  $c\simeq 0.2274$ is another constant found by numerical integration.
It may be readily checked that expressions (\ref{Znu1}) and
(\ref{ZD1}) differ from (\ref{Znu}) and (\ref{ZD}) only by a UV-finite renormalization
of the parameters $\nu$, $u_1$ and  $u_2$.

In the NP scheme, in contrast with the ${\rm \overline{MS}}$ scheme, the renormalized
Green functions have an analytic dependence on the set of parameters $\varepsilon$ and $\Delta$,
i.e. they do not have factors of the type $a\varepsilon+b\Delta$ in denominators. This is in
accord with the general ideas of the theory of analytic renormalization \cite{Speer}.

In the constants $Z$ of the ${\rm \overline{MS}}$ scheme with a fixed value of
$\zeta\equiv\Delta/\varepsilon=const$ the dependence on $\eps$ is present only in the
form of poles $1/\varepsilon$, $1/\varepsilon^2$ etc. Contrary to this, in the constants
$Z$ of the NP scheme regular terms $\sim 1,\,\varepsilon,\,\varepsilon^2$ etc. are added
to the poles in $\eps$. For calculation of the RG functions and the correction exponents
$\omega$ in Sec. \ref{sec:RGE} on rays (\ref{dz}) with $\zeta=\Delta/\varepsilon=const$
to the order $\varepsilon^2$ only terms of order $1/\varepsilon$ and $1$ are required
in the one-loop contributions $\sim u$ to $Z$, whereas in the two-loop contributions
$\sim u^2$ only terms of order $1/\varepsilon^2$ are $1/\varepsilon$ are needed. Expressions
(\ref{Znu1}) and (\ref{ZD1}) are quoted just with this accuracy.

\section{\label{sec:RGE}Renormalization-group  representation}

The use of renormalized parameters as such does not solve the main
problem of large expansion parameter growing with the Reynolds number. It is, however,
a necessary step towards the use of the method of the renormalization group which allows to solve
the problem by effective resummation of the perturbation theory. We shall consider
as an example the equal-time pair correlation function
\begin{eqnarray}
\label{Gr}
\langle \varphi_i(t,{\bf x}) \varphi_j (t,{\bf x'})\rangle \equiv
G_{ij}({\bf r})\,,\qquad {\bf r}\equiv {\bf x}-{\bf x'}\,,
\end{eqnarray}
which is the most interesting quantity for us in the following.
The Fourier transform of this function may be written as
\begin{equation}
\label{Gp}
G_{ij}({\bf p})= P_{ij}({\bf p})G(p),
\end{equation}
where $ P_{ij}({\bf p})$ is the transverse projection operator and $p\equiv | \bf p|$. Dimensional arguments
lead to the following representation of the scalar function $G(p)$ from Eq. (\ref{Gp}):
\begin{equation}
G(p) =  \nu^2p^{-d+2} R(s,g_1,g_2), \quad s={ \mu\over p}\,,
\label{dimG}
\end{equation}
where $R$ is a dimensionless function of dimensionless arguments. We want to calculate
$G(p)$ in the inertial range of the wave number $p$. Since in the present model (\ref{nakach})
the external scale of turbulence has been put equal to infinity, this corresponds to the
region $s= \mu/p\gg 1$. The perturbation expansion of $G(p)$ contains powers of the
parameter $s$ whose exponents grow without limit, due to which it is ill-suited for finding
the sought asymptotic behavior $s\to \infty$. We shall briefly remind solution of this problem within
the method of RG.

Since the fields $\Phi=\{\varphi,\,\varphi'\}$ in the present problem are not renormalized, the renormalized
functions $W^{R}$ differ from the unrenormalized ones $W=\langle\Phi\dots\Phi\rangle$ only by the
choice of variables and the form of perturbation expansion ($g_1$ and $g_2$ instead of $g_{10}$ and $g_{20}$), and
we may write:
\[
\label{WR}
W^{R}(g_1,g_2,\nu,\mu,\dots) = W (g_{10},g_{20},\nu_0,\dots)\,.
\]
Here, $e_0\equiv \{\nu_0, g_{10}, g_{20}\}$ is the set of all bare parameters,
whereas
$e\equiv \{\nu, g_{1}, g_{2}\}$ are their renormalized analogs, and the
ellipsis stands for the arguments not affected by renormalization like the coordinates, times etc.
The unrenormalized functions $W$ do not depend on $\mu$, while the renormalized functions $W^R$ do
because of the introduction of $\mu$ in renormalization relations (\ref{ZZ}).
The independence of $\mu$ of the functions $W$ is expressed by the
equation $\Dm W=0$. Here, and henceforth,  $\Dm\equiv \mu\partial_{\mu}$ with
fixed bare parameters $e_0$. The equation $\Dm W=0$ written in terms of the renormalized
functions $W^R=W$ and their arguments $e, \mu$ is
the basic RG equation
\begin{equation}
\Dm W^{R} (g,\nu,\mu,\dots)={\cal D}_{RG}W^{R} (g,\nu,\mu,\dots) = 0, \label{RGE}
\end{equation}
where
${\cal D}_{RG}$ stands for the operation $\Dm$ expressed in terms of the
renormalized variables:
\begin{equation}
\label{DRG}
{\cal
D}_{RG}\equiv {\cal D}_{\mu} +
\beta_1\partial_{g_1}+\beta_2\partial_{g_2}
-\gamma_{\nu}{\cal D}_{\nu}\,,
\end{equation}
where ${\cal D}_{x}\equiv x\partial_{x}$ for any variable $x$.
The RG coefficient functions (the anomalous dimensions
$\gamma$ and the $\beta$ functions) in Eq. (\ref{DRG}) are defined as
\begin{align}\label{gnu}
\gamma_a &\equiv \Dm \ln Z_{a}&a&\equiv \{\nu, g_1, g_2, D_2\}\nonumber\\
\beta_i &\equiv \Dm g_i,
,  &i&= 1, 2.
\end{align}
The term with ${\cal D}_{\nu}$ in Eq. (\ref{RGE}) is written with the account of
renormalization relation (\ref{ZZ}) for  $\nu$ and definition $\gamma_\nu$
(\ref{gnu}). From Eq.
(\ref{gnu}) and renormalization relations (\ref{ZZ}) it follows
\begin{align}
\label{b1}\beta_1(g_1,g_2)&
=
g_1\left[-2\varepsilon-\gamma_{g_1}(g_1,g_2)\right],\\
\label{b2} \beta_2(g_1,g_2) &=
g_2\left[2\Delta-\gamma_{g_2}(g_1,g_2)\right],
\\
\nonumber
\gamma_{g_1}&=-3\gamma_\nu,\quad
\gamma_{g_2}=\gamma_{D_2}-3\gamma_\nu.
\end{align}
We are interested in the IR asymptotics of small wave vectors
$\bf p$ and frequencies $\omega$ of the renormalized functions $W^R$ or, equivalently,
large relative distances and time differences in the ($t$, ${\bf x}$) representation
[in static objects like (\ref{Gr}) - (\ref{dimG}) dependence on $t$ or $\omega$ is absent].
It is determined by the IR-stable fixed point $g_*$, at which $\beta(g_*)=0$ for all $\beta$
functions. The fixed point $g_*$ is IR stable, if real parts of all eigenvalues of the
matrix $\omega_{ij}\equiv \partial\beta_i/\partial g_j\big|_{g=g_{*}}$ are strictly positive
(see, e.g. Refs. \cite{Zinn,Kniga}). Below it will be shown that in our model
(\ref{action3}) the system of two $\beta$ functions (\ref{b1}) and (\ref{b2})
in the region of our interest $\varepsilon>0,\, \Delta>0$ has an IR-stable fixed point
$g_*=\{g_{1*}, g_{2*}\}$ with $g_{1*}\neq 0, \,g_{2*}\neq 0$.

In its presence it follows from the RG equations (\ref{RGE}) that (see, e.g..
Refs. \cite{Kniga,turbo}) the sought asymptotics $W^R\big|_{IR}$ of the Green function
$W^R$ has the following property of ''IR scaling'' [in the
($t$, ${\bf x}$) representation]
\begin{align}\label{scal}
W^R\big|_{IR}(\lambda^{-\Delta_\omega} t,\lambda^{-1}{\bf x})&=
\lambda^{\Delta_W}W^R\big|_{IR}(t,{\bf x}),\nonumber\\
\Delta_W&=\sum_\Phi \Delta_\Phi ,
\end{align}
where ${\bf x}$ is the set of all coordinate variables and  $t$ all times, whereas
$\lambda>0$ is an arbitrary stretching parameter. Summation in expression
(\ref{scal}) for $\Delta_W$ goes over all fields $\Phi= \{\varphi, \varphi'\}$
entering the function $W^R$. In Eq.
(\ref{scal}) only those arguments of the function $W^R$ are explicitly shown which are
stretched under given scale transformation.

The quantities $\Delta_\omega$ and $\Delta_\Phi$ in Eq. (\ref{scal}) are critical dimensions
of the frequency $\omega$ and the fields $\Phi= \{\varphi,
\varphi'\}$. They are all unambiguously (see, e.g..
Refs. \cite{Kniga,turbo}) expressed through the quantity $\gamma_{\nu}^*\equiv \gamma_\nu(g_*)$
-- the value of the RG function $\gamma_\nu(g)$, defined in Eq. (\ref{gnu}),
at the fixed point:
\begin{align}\label{Delta}
\Delta_\varphi&=1 - \gamma_{\nu}^*\,, &\Delta_{\varphi'}& = d -
\Delta_\varphi\,,\nonumber\\
\Delta_\omega &= 2 -
\gamma_{\nu}^*\,,&\gamma_{\nu}^*&\equiv \gamma_\nu(g_*).
\end{align}
At the fixed point with
$g_{1*}\neq 0$ and $ g_{2*}\neq 0$ the values
$\gamma_a^*\equiv \gamma_a(g_*)$ of RG functions (\ref{gnu}) are readily
found from the definition the fixed point $\beta_1(g_*)=\beta_2(g_*)=0$ together with relations
(\ref{b1}) and  (\ref{b2}):
$\gamma_{g_1}^*=-2\varepsilon,\,\, \gamma_{g_2}^*=2\Delta,\,\,
\gamma_\nu^*=2
\varepsilon/3,\,\,\gamma_{D_2}^*=2\Delta+2\varepsilon $.
Substitution of $\gamma_\nu^*=2 \varepsilon/3$ in Eq. (\ref{Delta})
leads to formulas (\ref{razm}) and their corollaries (\ref{kolm}) for
$\varepsilon=2$. Thus, in two-charge model
(\ref{action3}) with the local renormalization \cite{Nalimov}
the critical dimensions of the velocity field $\varphi$ and frequency $\omega$
at the real value $\varepsilon=2$ retain their Kolmogorov values contrary to
the conjecture of the author of Ref. \cite{Ronis}.

Consider again function (\ref{dimG}). It is a particular case of the function
$W^R$ and satisfies RG equation (\ref{RGE}): ${\cal D}_{RG} G=0$.
A representation of the solution of Eq. (\ref{RGE}) for $G(p)$ convenient for
the asymptotic analysis at $p\rightarrow 0$ may be obtained with the aid of
invariant variables $\bar e = \bar e(s,e)$ corresponding to the complete set
of renormalized parameters $e\equiv\{\nu,g_1,g_2\}$. They are defined as solutions
of the RG equations ${\cal D}_{RG} \bar e =0$ with the operator ${\cal D}_{RG}$ from Eq. (\ref{DRG})
and the normalization conditions $\bar e = e$ at $s=1$. In terms of the invariant
variables the solution of the RG equation (\ref{RGE}) for $G(p)$ may be represented as
\begin{equation}
G(p) =  \nu^2p^{2-d} R(s,g_1,g_2)=\bar\nu^2 p^{2-d}R(1,\bar
g_1,\bar g_2). \label{dimG1}
\end{equation}
The right-hand side of (\ref{dimG1}) depends on $s$ through the invariant
variables $\bar e(s,e)$ only, whose asymptotic behavior in the limit
$s\rightarrow \infty$ -- determined by the IR stable fixed point
(see below) -- is simple: the invariant charges $\bar g_1$ and $\bar g_2$ tend to fixed values
$g_{1*}=O(\varepsilon)$ and $g_{2*}=O(\varepsilon)$, whereas the invariant viscosity has simple
powerlike asymptotics. It may be conveniently determined by expressing the invariant variables
$\bar e = (\bar \nu, \bar g_1, \bar g_2)$
in terms of the bare variables $e_0=(\nu_0, g_{10}, g_{20})$ and the wave number $p$.
According to definition, the bare variables $e_0$ as well as
the invariant variables $\bar e$ satisfy the equation
${\cal D}_{RG} e_0 =\Dm e_0= 0$. The connection between the two sets of parameters is determined
by the relations
\begin{align}
 \nu_0&=\bar \nu Z_{\nu}(\bar g), & g_{10}&=\bar g_1
p ^{2\eps}Z_{g_1}(\bar g), \nonumber\\ g_{20}&=\bar g_2 p
^{-2\Delta}Z_{g_2}(\bar g)\,,&& \label{119}
\end{align}
valid because
both sides in each of them satisfy the RG equation, and because they
at $s\equiv \mu/p =1$ coincide with relations (\ref{ZZ}) owing
to the normalization conditions. Using the connection between renormalization constants $Z_gZ_\nu^3=1$
indicated in Eq. (\ref{ZZ}) and excluding these constants from the first two relations in
(\ref{119}) we find
$g_{10}\nu_0^3=D_{10}={\bar g_1}p^{2\varepsilon}\,{\bar
\nu}^{\,3}$, and from here
\[
\bar \nu=(D_{10}p^{-2\varepsilon}/\, {\bar g}_{1})^{1/3}\,,
%\label{4}
\]
which for the sought asymptotics
$s\rightarrow \infty$ with the account of $\bar
g_1 \rightarrow g_{1*}$ yields
\begin{eqnarray}
\bar\nu \to \bar \nu_*=( D_{10} / g_{1*}
)^{1/3}p^{-2\varepsilon/3}, \quad \quad s\rightarrow \infty .
\end{eqnarray}
Substituting this result in Eq. (\ref{dimG1}) we obtain
\begin{equation}
G(p) \simeq (D_{10} / g_{1*} )^{2/3} p^{2-d-4\varepsilon/3}R(1,
g_{*}), \quad \quad s\rightarrow \infty . \label{asymp}
\end{equation}
This relation will be used in Sec. \ref{sec:Q}.

Let us make a remark about relations (\ref{119}). According to renormalization
relations (\ref{ZZ}), condition $D_{20}\sim g_{20}=0$
[see the text following Eq. (\ref{nakach2})] imposes the constraint
$Z_{D_2}(g)\sim Z_{g_2}(g)=0$ on the renormalized charges
$g=\{g_1,\,g_2\}$. From the last relation in Eq. (\ref{119}) it follows that
the invariant charges $\bar g=\bar g(s,g)$ for any value of the variable
$s\equiv \mu/p$ lie on the same constraining surface $Z_{g_2}=0$ as the initial data $\bar g |_{s=1}=g$.
Therefore, the limit values
$g_*=\lim\limits_{s\rightarrow\infty}\bar g(s,g)$ lie on the same surface  $Z_{g_2}=0$,
i.e. the condition $D_{20}\sim g_{20}=0$ is compatible with the RG analysis.

All said above is valid for any subtraction scheme, only the explicit
form of the RG functions $\gamma_a$ in Eqs.
(\ref{gnu}) and (\ref{b1}) depends on the choice of the scheme.
We shall first quote results of the two-loop calculation in the ${\rm \overline{MS}}$ scheme
(Sec. \ref{sec:local}) and then briefly discuss modification of formulas in the NP scheme. As said
before, no physically significant results depend on the choice of the scheme.

In the MS and ${\rm \overline{MS}}$ schemes all RG functions $\gamma_a$
are independent of $\varepsilon$. In model (\ref{action3}) they depend only
on charges and the parameter $\zeta=\Delta/\varepsilon$.
The two-loop expressions for the constants $Z_a$ in Eq. (\ref{gnu}) are given by Eqs.
(\ref{Znu}) and (\ref{ZD}). In calculation of the quantities $\gamma_a = \Dm
\ln Z_{a}$ from Eq. (\ref{gnu}) the operation $\Dm$ may be replaced by  ${\cal
D}_{RG}$ from Eq. (\ref{RGE}) and the contributions with ${\cal D}_\mu$
and ${\cal D}_\nu$ omitted, since the quantities $Z_a$ do not depend on $\mu$ and $\nu$.
Such a calculation yields
\begin{equation} \label{gnu2}
\gamma_{\nu}=2\,({u_1}+{u_2})+\frac{2(4\zeta+3)u_1^2}{2+\zeta}
+2(5\zeta+3)u_1u_2-4R\,({u_1}+{u_2})^2+\,...\,,
\end{equation}
\begin{multline} \label{gD}
\gamma_{D_2}=\frac{2\,({u_1}+{u_2})^2}{u_2}-\frac{\zeta(13+19\zeta)u_1^3}
{(2+\zeta)u_2}\\
+\frac{2(34\zeta+19+6\zeta^2)u_1^2}{2+\zeta}
-6u_1^2+(13+31\zeta)u_1u_2\\
+\frac{4(1-R)\,({u_1}+{u_2})^3}{u_2}+\,...\,\,.
\end{multline}
Let us remind that $u_1\sim g_1$ and $u_2\sim g_2$ are charges with a more
convenient normalization (\ref{alpha2}), while the ellipsis stands for corrections of order
$O(u^3)$.

Substituting quantities (\ref{gnu2}) and (\ref{gD}) in Eq. (\ref{b1}) we obtain
expressions for the $\beta$ functions in the two-loop approximation. Then from the
conditions $\beta_1(g_*)=\beta_2(g_*)=0$ coordinates of the fixed points
$g_*\sim u_*$ may be found. In the framework of the
$\varepsilon$ expansion there are three fixed points \cite{Nalimov}:
1) the trivial fixed point $u_{1*}=0,\,\,u_{2*}=0$;
2) the "kinetic" fixed point $u_{1*}=0,\,\,u_{2*}\neq 0$; and
3) the "Kolmogorov" fixed point $u_{1*}\neq 0,\,\,u_{2*}\neq 0$.
In the region
$\varepsilon>0,\,\Delta>0$ of interest for us only the Kolmogorov fixed point is
IR stable, for which in the one-loop approximation
\begin{equation}
u_{1*}+u_{2*}=\frac{\varepsilon}{3}+O(\varepsilon^2),\quad
u_{2*}=\frac{\varepsilon}{9(1+\zeta)}+O(\varepsilon^2). \label{x}
\end{equation}
From relations (\ref{gnu2}) and (\ref{gD}) two-loop
contributions $\sim\varepsilon^2$ to Eq. (\ref{x}) may be found. We do not quote
them, because coordinates of a fixed point $u_*\sim g_*$ do not have direct
physical meaning and do depend on the choice of the subtraction scheme. Objective
quantities independent of the subtraction scheme are the eigenvalues of the matrix
$\omega_{ij}=\partial\beta_i/\partial g_j\big|_{g=g_{*}}$. In our problem the $\omega$
matrix is a $2\times2$  matrix, whose two eigenvalues $\omega_{\pm}$ in the two-loop
approximation at the Kolmogorov fixed point are
\begin{equation}\label{omegi}
\omega_{\pm}=\left(\zeta+\frac{4}{3}\pm\frac{\sqrt{9\zeta^2-12\zeta-8}}{3}\right)\,\epsilon+
\frac{2}{9}\left\{-3-2R-3\zeta\pm\frac{\left[4 (1+3\zeta) R-6-12 \zeta-9
\zeta^2\right]}{\sqrt{9 \zeta^2-12 \zeta-8}}\right\}\,\epsilon^2.
\end{equation}
We quote also for reference the relatively simple expressions for the trace and determinant
of the $\omega$ matrix, through which the eigenvalues $\omega_{\pm}$ are unambiguously expressed:
\begin{align}\label{tr}
\text{Tr}\,\omega&=\omega_+ +\omega_- =
\frac{2}{3}(3\zeta+4)\epsilon-\frac{4}{9}(3\zeta+3+2R)\epsilon^2,\\\label{det}
\det\,\omega&=\omega_+ \omega_-
=\frac{4}{3}(3\zeta+2)\epsilon^2-\frac{4}{9}(2R+1)(3\zeta+2)\epsilon^3.
\end{align}
The one-loop contributions $\sim \varepsilon$ in Eqs.
(\ref{x}) - (\ref{tr}) and  $\sim\varepsilon^2$ in Eq. (\ref{det})
were obtained earlier in Ref. \cite{Nalimov}. In the one-loop approximation
this fixed point $g_*$ is IR stable in the sector
$\varepsilon>0,\,\zeta>-2/3$ in the ($\varepsilon,\,\Delta$) plane.
When $\varepsilon>0$ and $\zeta<-2/3$ both eigenvalues
(\ref{omegi}) are real and have different signs [most easily this may be seen
from the one-loop contribution in Eqs. (\ref{tr}) and (\ref{det})]. With
growth of $\zeta$ upon intersection of the borderline $\zeta_0=-2/3$ both 
eigenvalues become positive and then, upon reaching the next borderline $2(1-\sqrt{3})/3\simeq - 0.488$,
the argument of the root in Eq. (\ref{omegi}) becomes negative, i.e. the fixed point
becomes an IR-attractive focus with $\omega_{\pm}=a\pm ib$ with $a>0$.
It remains such until the next borderline $2(1+\sqrt{3})/3\simeq 1.821$ is reached,
upon passing which the root argument in Eq. (\ref{omegi}) becomes positive again and both
eigenvalues $\omega_{\pm}$ real and positive. For our "physical" ray
$\zeta=1/4\,\,(d=3)$ the fixed point $g_*$ is an IR-attractive focus.

What was said above refers to the one-loop approximation. The account of the two-loop
corrections in Eqs. (\ref{omegi}) - (\ref{det}) leads to a deformation of the borderlines
of the region of IR stability, but the "physical" segment of ray (\ref{dz})
with $\zeta=1/4$,  $0<\varepsilon\leq 2$ still remains in this region.

Consider now renormalization in the NP scheme, in which the constants $Z$
are determined from the normalization condition (\ref{Rato3}). The two-loop
expressions for the constants $Z$ with the necessary accuracy [see the text
following Eq. (\ref{ZD1})] are given in Eqs. (\ref{Znu1}) and (\ref{ZD1}). With the use of them
together with definitions (\ref{gnu}) we obtain the following expressions for $\gamma_a$:
\begin{equation}
\label{gnu1} \gamma_\nu=  \left( 2+3\,\Delta- {c}\,\epsilon\,
\right) u_{{1}}+ \left(2+ {c}\,\Delta\,+3 \,\Delta\right)
u_{{2}}
-4\, \left( u_{{1}}+u_{{2}} \right) ^{2}{( 2R+1)}\,,
\end{equation}
\begin{multline} \label{gD1}
\gamma_{D_2}={ \frac {{u_{{1}}}^{2} \Big[
2+(7-{c})\,\Delta\,+(4-2\,{c})\, \epsilon\, \Big] }{u_{{2}}}}\\
 +2\, \Big[ 2+5\, \Delta+(2-\,{c})\epsilon\, \Big]
u_{{1}}+ \Big[2+ \, (3+{c})\Delta\, \Big] u_{{2}}\\
-4\,{\frac {
\left( u_{{1}}+u_{{2}} \right) ^{3}{( 2R+1)}}{u_{{2}}}}\,,
\end{multline}
where the notation is the same as in Eqs. \ref{Znu1}) and
(\ref{ZD1}). The RG functions (\ref{gnu1}) and (\ref{gD1}), contrary to their
analogs in the ${\rm \overline{MS}}$ scheme, do not contain factors like
$\zeta+const$  in denominators, i.e. they are analytic in the pair of parameters
$\varepsilon,\,\Delta$, which is a consequence of similar analyticity of the renormalized
Green functions.
Coordinates of fixed points $u_*\sim g_*$ obtained from Eqs. (\ref{gnu1}) and (\ref{gD1})
in the one-loop approximation keep the form of Eq. (\ref{x}), but the two-loop contributions
(which we do not quote) differ from analogous contributions in the  ${\rm
\overline{MS}}$ scheme. The eigenvalues $\omega_{\pm}$ of the matrix
$\omega$, however, remain exactly the same as in the ${\rm \overline{MS}}$ scheme, because these
quantities do not depend on the subtraction scheme.

In conclusion, we note that in an attempt to use the NP scheme
(\ref{Rato1}) in the model \cite{Ronis} with nonlocal renormalization, the inconsistency of this model
in terms of the RG functions $\gamma_a$ would show in the form of poles
$1/\varepsilon$ in the two-loop contributions.

\section{\label{sec:Q}Skewness factor and Kolmogorov constant}

The exponent of the power of the wave number in Eq. (\ref{asymp}) is determined exactly and does not
have corrections in the form of higher powers of $\eps$. At the physical value
$\eps=2$ this exponent assumes the Kolmogorov value.
To find the Kolmogorov constant, the amplitude of this
function has to be calculated, which, however, can be done only
approximately, because the corresponding $\eps$ expansion does not
terminate. In calculation of the amplitude, apart from technical
difficulties at two-loop order, a principal problem arises as
well. It is connected with the necessity to express the answer for
$G(k)$ in terms of the energy injection rate $\E$ instead of the
parameter $D_{10}$ of the forcing correlation (\ref{nakach2}).
The connection between $D_{10}$ and $\E$ is determined by an exact relation expressing
$\E$ in terms of the function $d_f(k)$ in the correlation function (\ref{1.2})
\begin{equation}
\E = \frac{(d-1)}{2 (2\pi)^{d}} \,\int\! d {\bf k} \, d_f(k).
\label{bal}
\end{equation}
Substituting here function (\ref{nakach2}) with $D_{20}=0$ [see the text
following Eq. (\ref{nakach2})] and introducing the UV cutoff  $k\leq \Lambda =
(\E/\nu_0^3)^{1/4}$ (the inverse dissipation length), we obtain the following
connection between the parameters $\E$ and $D_{10}$
\begin{eqnarray}
D_{10} = \frac{4(2-\eps)\,\Lambda^{2\eps-4}} {\overline
S_{d}(d-1)}\,\, \E \,. \label{2.74}
\end{eqnarray}
Idealized injection by infinitely large eddies corresponds to
$d_{f}(k) \propto \delta({\bf k})$. More precisely, according to Eq.
(\ref{bal})
\begin{equation}
d_f(k)= {2 (2\pi)^{d}\, \E\, \delta({\bf k}) \over d-1}. \label{2.75}
\end{equation}
In view of the relation
\[
\delta({\bf k}) = \lim_{\eps\to 2}\,(2\pi)^{-d}\int\! d{\bf
x}(\Lambda x)^{2\varepsilon-4}\exp(i{\bf k}{\bf x})\\
=S_{d}^{-1}
k^{-d} \lim_{\eps\to 2} \left[ (4-2\eps) (k/\Lambda)^{4-2\eps}
\right],
%\label{2.76}
\]
the powerlike injection with
$d_f=D_{10}k^{4-d-2\varepsilon}$ and the amplitude
$D_{10}$ from Eq. (\ref{2.74}) in the limit $\varepsilon\rightarrow 2$ from the
the region $0<\varepsilon<2$ gives rise to the $\delta$ sequence
(\ref{2.75}).

Relation (\ref{2.74}) reveals that at fixed
$\E$ the quantity $D_{10}$ depends on $\eps$ and it is necessary to take this
dependence into account in the construction of the $\eps$ expansion for the
Kolmogorov constant.
On the other hand it shows that the quantity $R(1,g_{*})$ from (\ref{asymp})
must have a singularity of the form $(2-\eps)^{-2/3}$ in the limit $\eps\to 2$:
only in this case the Kolmogorov constant in the model with the injection
$d_f=D_{10}k^{4-d-2\varepsilon}$ and the amplitude $D_{10}$ from Eq.
(\ref{2.74}) shall have a finite value in the limit $\varepsilon\rightarrow 2 $.
The measurable experimental Kolmogorov constant $C_K$ in terms of the model with
such pumping corresponds to the limiting value
$\varepsilon=2$, and we want to define its generalization $C_K(\varepsilon)$ for the whole
interval $0\leq \varepsilon \leq 2$. Obviously, such a generalization cannot be done
unambiguously, because it is not possible to define unambiguous dependence
of the parameter $D_{10}$ in Eq. (\ref{2.74}) on $\varepsilon$ at a fixed value of $\E$.

Let us explain this in more detail. When deriving relation (\ref{2.74}) we assumed
that integral (\ref{bal}) for the injection $d_f=D_{10}k^{4-d-2\varepsilon}$ has an
upper cutoff equal to the inverse dissipative length $ \Lambda = (\E/\nu_0^3)^{1/4}$.
Such a cutoff is natural, but at the same time only orders of magnitude may be discussed, of course,
not the exact values. Therefore, there is nothing to prevent to replace in Eq.
(\ref{2.74}) the cutoff parameter $\Lambda$ by $a\Lambda$ with a coefficient
$a$ of the order of unity, which yields
the extra factor $a^{2\varepsilon-4}$ on the right-hand side of Eq. (\ref{2.74}).
This factor tends to unity at $\varepsilon\rightarrow 2$,
hence it does not affect the the physical (real) value of the Kolmogorov constant
$C_K(\varepsilon=2)$, but it does affect coefficients of the hypothetical
$\varepsilon$ expansion of the function $C_K(\varepsilon)$.
Generalizing these observations it may be stated that the physical content of the theory
is not changed, if to the right-hand side of Eq. (\ref{2.74}) an extra factor $F(\eps)$ with $F(2)=1$
is added. In Ref.~\cite{JETP} (see also~\cite{Kniga,UFN,turbo}) relation (\ref{2.74})
without the extra factor $F(\eps)$ was regarded as the definition of the quantity $D_{10}$.
Other approaches to the definition of the function $C_K(\varepsilon)$ and its $\varepsilon$
expansion \cite{48} - \cite{Jap} may be reduced
to the introduction of a particular function
$F(\eps)$ with $F(2)=1$ on the right-hand side of relation (\ref{2.74}).

Thus, $\eps$ expansion of the Kolmogorov constant in the model with the powerlike
injection is not defined unambiguously. However, physical quantities independent of the
amplitude $D_{10}$ (universal quantities) do have a well-defined $\eps$ expansion.
The skewness factor
\begin{equation}
{\cal S} \equiv S_{3}/S_{2}^{3/2}, \label{sviaz_2}
\end{equation}
is an example of such a quantity. In Eq. (\ref{sviaz_2})
$S_n$ are structure functions defined by relations
\begin{equation}
S_{n} (r) \equiv \big\langle [ \varphi_{r} (t,{\bf x}+{\bf r}) -
\varphi_{r} (t,{\bf x})]^{n} \big\rangle, \  \varphi_{r}\equiv
{(\varphi_{i}\cdot r_{i})\over |{\bf r}|}. \label{struc}
\end{equation}
According to Kolmogorov theory, the structure function $S_{2}(r)$ in the inertial range
is of the form
\begin{equation}
S_{2}(r) = C_K  \E^{2/3}r^{2/3},
 \label{CK}
\end{equation}
where $C_K $ is the Kolmogorov constant with a simple connection with
the Kolmogorov constant of the energy spectrum \cite{Monin}.
Since the experimental evidence for anomalous scaling [i.e. deviation of the power of $r$ from the Kolmogorov
value ${2\over 3}$ in Eq. (\ref{CK}) in the inertial range] in the structure function $S_{2}(r)$
is still controversial and in any case this deviation is small \cite{Barenblatt99}, we shall use the
Kolmogorov asymptotic expression (\ref{CK}) in the following analysis.

The structure function $S_{3}(r)$ may be found exactly in the inertial range \cite{Monin}:
\begin{equation}
S_{3}(r) = -\frac{12}{d(d+2)}\,\E\, r,
\label{S}
\end{equation}
which allows -- with the account of (\ref{sviaz_2}) and (\ref{CK}) --
to relate the Kolmogorov constant and the skewness factor
\begin{equation}
C_K=\Big[-\frac{12}{d(d+2)\cal S}\Big]^{2/3}. \label{sviaz}
\end{equation}
Of the three quantities $S_{2}(r)$,  $S_{3}(r)$ and $\cal S$ only $\cal S$ has
a unique $\varepsilon$ expansion. Thus, relation (\ref{sviaz}) (valid only for the physical value$\varepsilon=2$)
might be used to determine $C_K$ using the calculated value ${\cal S} (\varepsilon=2)$.

To find the RG representation of skewness factor (\ref{sviaz_2}) it is necessary to have
RG representations of the functions $S_{2}(r)$ and $S_{3}(r)$.
The function $S_{2}(r)$ is connected with the Fourier transform of the
pair correlation function $G(k)$ by the relation
\begin{equation}
S_{2}(r) = 2\int \frac{d{\bf k}}{(2\pi)^{d}} \,G(k)\,
\left[1-{({\bf k}\cdot{\bf r})^{2}\over (kr)^{2}}\right] \left\{1- \exp
\left[{\rm i} ({\bf k}\cdot{\bf r})\right]\right\}, \label{atas}
\end{equation}
therefore its RG representation may be found on the basis of
RG representation (\ref{asymp}). An analogous RG representation in the inertial interval
may be written for the function $S_{3}(r)$. It is more convenient, however, to use the
following exact result, an analog of expression (\ref{S}):
\begin{equation}
S_3(r)= -\frac{3(d-1) \, \Gamma(2-\eps) \, (r/2)^{2\eps-3}D_{10}}
{(4\pi)^{d/2} \, \Gamma(d/2+\eps)}\,. \label{S3}
\end{equation}
This relation is a manifest demonstration that the amplitude of the structure function,
expressed in terms of $D_{10}$, has a singularity at $\varepsilon\rightarrow 2$, in this case
it is $\sim (2-\varepsilon)^{-1}$. On substitution of Eq. (\ref{2.74}) in Eq. (\ref{S3}) this
singularity cancels the corresponding zero on the right-hand side of Eq. (\ref{2.74}), leading for
$S_{3}(r)$ to an expression finite at
$\varepsilon=2$ and coinciding with Eq. (\ref{S}).

Relations (\ref{asymp}), (\ref{atas}) and (\ref{S3}) might serve as the
basis for construction of the $\varepsilon$ expansion of the skewness factor
(\ref{sviaz_2}). However, an additional difficulty arises on this way. The point is
that the powerlike dependence $S_{2}(r)\sim r^{2-2\varepsilon/3}$, determined from Eqs.
(\ref{asymp}) and (\ref{atas}), is only valid when
$\varepsilon>3/2$, because for $\varepsilon<3/2$ integral
(\ref{atas}) diverges at $k\rightarrow \infty$ [this means that
the main contribution to $S_{2}(r)$ in this case is given by the term
$\big\langle \varphi_{r}^{2} (t,{\bf x}) \big\rangle$ independent of $r$].
However, the derivative $r\partial_r S_{2}(r)$ is free from this flaw, because,
according to Eq. (\ref{atas}),
\begin{equation}
 r\partial_r S_{2}(r) = 2 \int \frac{d{\bf k}}{(2\pi)^{d}}\,
G(k)\,
\left[1-{({\bf k}\cdot{\bf r})^{2}\over (kr)^{2}}\right]\, ({\bf
k}\cdot{\bf r}) \,\sin ({\bf k}\cdot{\bf r}). \label{atas1}
\end{equation}
Integral (\ref{atas1}) is convergent for all $0<\varepsilon <2$.
On the other hand, at the physical value $\varepsilon=2$ the amplitudes in
$S_{2}(r)$ and  $r\partial_rS_{2}(r)$ differ by a trivial factor ${2\over 3}$,
therefore in Refs. \cite{Adzhemyan03a,slovac,Adzhemyan03b} for the construction of
the $\varepsilon$ expansion the following analog of the skewness factor was used
\begin{equation}
Q(\eps)\equiv\,{  r\partial_r S_{2}(r) \over |S_{3}(r)|^{2/3}}= {
r\partial_r S_{2}(r) \over (-S_{3}(r))^{2/3}}. \label{Q}
\end{equation}
The Kolmogorov constant and the skewness factor are expressed through the value
$Q(\varepsilon=2)$ according to Eqs. (\ref{sviaz_2}), (\ref{CK}) and
(\ref{S}) by the relations
\begin{equation}
C_{K}= \frac{3Q(2)}{2}\st{\frac{12}{d(d+2)}}^{2/3}\!\!\!, \
{\cal S} = - \left[\frac{2}{3Q(2)}\right]^{3/2}\!\!\!. \label{trud}
\end{equation}
Quantity (\ref{Q}) may be calculated both in the double
($\varepsilon$, $\Delta$) expansion and in the usual $\varepsilon$ expansion.
In the former case the corresponding expansion is obtained on the basis of relations
(\ref{asymp}), (\ref{S3}) and (\ref{atas1}) in the form:
\begin{equation}
Q(\eps,\zeta)=\eps^{1/3}\sum_{k=0}^\infty \Psi_k(\zeta)\eps^k .
\label{Qepsdz}
\end{equation}
The usual $\varepsilon$ expansion of the quantity $Q$ for dimensions
$d>2$ has been obtained in Ref. \cite{Adzhemyan03a}:
\begin{equation}
Q(\eps, d)=\eps^{1/3}\sum_{k=0}^\infty Q_k(d)\eps^k .
\label{Qeps}
\end{equation}
The connection between expansions (\ref{Qepsdz}) and (\ref{Qeps}) is revealed
by investigation of singularities of the coefficients $Q_k(d)$ in Eq.
(\ref{Qeps}) at $d\rightarrow 2$. An analysis of these singularities shows
that in the vicinity of $d-2=2\Delta=0$ these coefficients may be expressed in
a Laurent expansion
\begin{equation}
Q_k(d) = \sum_{l=0}^\infty q_{kl}\, \Delta^{l-k}.
 \label{Qq}
\end{equation}
Substitution of expression (\ref{Qq}) in Eq. (\ref{Qeps}) leads to the
representation
\begin{equation}
Q(\eps, d)=\eps^{1/3}\sum_{k=0}^\infty \sum_{l=0}^\infty
(\eps/\Delta)^k q_{kl} \, \Delta^l. \label{Qed}
\end{equation}
Changing variables in Eq. (\ref{Qed}) to  $\varepsilon$ and
$\zeta=\Delta/\varepsilon$, we arrive at expansion (\ref{Qepsdz}),
in which
\begin{equation}
\Psi_k(\zeta)=\sum_{l=0}^\infty
 q_{lk} \,\zeta^{k-l}. \label{psi}
\end{equation}
Relations (\ref{Qq}) and (\ref{psi}) show that the alternative
$\varepsilon$ expansions (\ref{Qepsdz}) and
(\ref{Qeps}) sum different infinite subsequences of double sum
(\ref{Qed}). In Ref. \cite{Adzhemyan03b} a procedure of improvement of the
$\eps$ expansion was proposed with the use of the mutually complementary information
about the quantity $Q$ contained in the partial sums of
expansions (\ref{Qepsdz}) and (\ref{Qeps})
\begin{equation}
Q^{(n)}_{\eps, \Delta} \equiv \eps^{1/3}\sum_{k=0}^{n-1}
\Psi_k(\zeta)\eps^k , \ Q^{(n)}_{\eps} \equiv
\eps^{1/3}\sum_{k=0}^{n-1} Q_k(d)\eps^k , \label{Qedn}
\end{equation}
where $n\geq 1$ is the number of loops.

\begin{figure}[!]
\includegraphics[width=9cm]{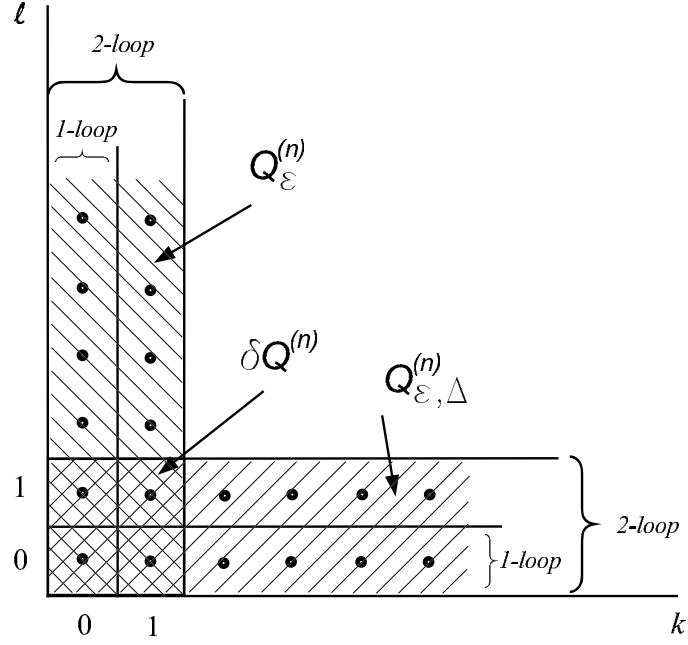}
\caption{\label{qed1}
Summations in the calculation of $Q^{(n)}_{eff}$ in Eq. ( \ref{Qeffn}).
Terms in the double sum (\ref{Qed})
taken into account in $Q^{(n)}_{\eps, \Delta}$ and $Q^{(n)}_{\eps}$,
correspond to the dashed
horizontal and and vertical stripes, respectively. The correction term
$\delta Q^{(n)}$ corresponds
to sum over the double-dashed square.}
\end{figure}

Terms in the double sum (\ref{Qed})
taken into account in $Q^{(n)}_{\eps, \Delta}$ and $Q^{(n)}_{\eps}$,
have been schematically plotted in Fig. \ref{qed1} in the form of dashed
horizontal and and vertical stripes, respectively.

All terms in the dashed area will be taken into account in the effective quantity
\begin{equation}
Q^{(n)}_{eff} = Q^{(n)}_{\eps} + Q^{(n)}_{\eps,\Delta} - \delta
Q^{(n)} ,
 \label{Qeffn}
\end{equation}
where
\[
\delta Q^{(n)} \equiv  \eps^{1/3}\sum_{k=0}^{n-1} \sum_{l=0}^{n-1}
(\eps/\Delta)^k q_{kl} \, \Delta^l
% \label{dQen}
\]
is a subtraction term necessary to avoid double counting of terms with
$k\leq n-1,\,l\leq n-1$ (the double-dashed square in Fig.\ref{qed1}).
It may be found by taking the corresponding number of terms from expansions
(\ref{Qq}) or (\ref{psi}). From the point of view of the usual
$\eps$ expansion (\ref{Qeps}) relation (\ref{Qeffn}) may be interpreted
as follows: in the $n-1$ first terms of the expansion the coefficients
$Q_k(d)$ from Eq. (\ref{Qeps}) are calculated exactly, but in all higher-order terms ($k\ge n$)
approximately with the account of $n-1$ first terms of their Laurent expansion (\ref{Qq}).

Our two-loop calculation of the $\eps,\Delta$ expansion of the quantity
$Q$ together with the two-loop calculation of Ref. \cite{Adzhemyan03a}
allowed to obtain an improved $\eps$ expansion of the quantity $Q$ at second order
of perturbation theory \cite{Adzhemyan03b}. For the Kolmogorov constant calculated
according to Eq. (\ref{trud}) for $d=3$ it led to the result quoted in Table \ref{tab:table1}.

\begin{table}
\caption{\label{tab:table1}
One and two-loop values of the Kolmogorov
constant in the usual $\eps$ expansion
($C_{\eps}$) and the double $\eps, \Delta$ expansion
($C_{\eps,\Delta}$); the contribution $C_{\delta}$ in Eq. (\ref{trud}) from the
correction $\delta Q^{(n)}$ in Eq. (\ref{Qeffn}),
and the value $C_{eff}$ from Eqs. (\ref{trud}), (\ref{Qeffn}).}
\begin{ruledtabular}
\begin{tabular}{lrrrr}
n&$C_{\eps}$&$C_{\eps,\Delta}$&$C_{\delta}$&$C_{eff}$\\
\hline
1 & 1.47 & 1.68 & 1.37 & 1.79 \\
2 & 3.02 & 3.57 & 4.22 & 2.37 \\
\end{tabular}
\end{ruledtabular}
\end{table}

In Table \ref{tab:table1}
%~\ref{tab:table1}
we have quoted for comparison the
values of the Kolmogorov constant calculated according to
Eq. (\ref{trud}) at first and second order of the usual $\eps$
expansion ($C_{\eps}$), the double $\eps, \Delta$ expansion
($C_{\eps,\Delta}$), the contribution $C_{\delta}$ in Eq. (\ref{trud})
from the correction $\delta Q^n$ in Eq. (\ref{Qeffn}) and the value
$C_{eff}$ obtained from relations (\ref{trud}) and (\ref{Qeffn}). In all the
cases quoted the recommended experimental value of the Kolmogorov
constant $C_{exp}=2.01$ \cite{Sreenivasan95} lies between the
values of the first and second approximation. However, the
difference between these values is rather significant both in the
$\eps$ expansion and in the ($\eps, \Delta$) expansion, let alone
the leading terms of the $\eps$ expansion of the latter. For the
improved $\eps$ expansion, i.e. for the quantity
$C_{eff}=C_{\eps}+C_{\eps,\Delta}-C_{\delta}$ calculated according
to Eqs. (\ref{Qeffn}) and (\ref{trud}), however, this difference is
about three times smaller leading to far better agreement with the
experimental data.

\section{\label{sec:conclusion}Conclusion}

In conclusion, we have presented a detailed comparison of
two different renormalization schemes for the stochastic Navier-Stokes
problem near two dimensions. By explicit two-loop calculation
we have shown that the nonlocal scheme of Ref. \cite{Ronis}
cannot consistently be carried out beyond the leading one-loop approximation.
On the contrary, our two-loop results confirm the consistency
of the local renormalization scheme of Ref. \cite{Nalimov} based on the
general principles of the theory of UV renormalization.

The detailed explicit two-loop analysis of different renormalization schemes presented
here is all the more important, because the inconsistent renormalization
of nonlocal terms in dynamic models continues to appear in the literature \cite{Ronis2,Chen2001}.

The correct choice of the renormalization scheme is vital for
a proper account of the effect on structure functions of the
additional singularities appearing
in the field-theoretic model in the limit $d\to 2$.
Using the consistent local renormalization scheme,
we have shown that a proper account of the ''nearest singularity''
in the coefficients of the $\eps$ expansion (\ref{Qeps}) leads to a
significant improvement
of the results of the two-loop RG calculation
at $d=3$. We have analysed the effect ot this procedure at other $d$ as well.
It turned out to reduce significantly the relative contribution
of the two-loop correction in the whole range considered
$\infty > d\geq 2.5$. At the same time this contribution remained large
at $d = 2$, which we think to be an effect of singularities at the next exceptional
dimension $d = 1$.

The proposed procedure of approximate summation of the
$\varepsilon$ expansion is, of course, applicable not only to the calculation of
$Q(\eps)$, but all universal quantities such as
dimensions of composite operators.

\acknowledgements
The authors thank N.V. Antonov for
discussions. The work was supported by the Nordic Grant for
Network Cooperation with the Baltic Countries and Northwest Russia
Nos.~FIN-6/2002 and FIN-20/2003. L.Ts.A. and M.V.K.
acknowledge the Department of Physical Sciences of the University
of Helsinki for kind hospitality.

\appendix*
\section{\label{sec:Lambda}$\Lambda$ renormalization and ($\eps$, $\Delta$) expansion above two dimensions}

As it was explained in Sec. \ref{sec:local}, in the two-charge
model (\ref{action3}) in some graphs the wave-vector integrals diverge at large
wave numbers. To regularize such integrals, it is necessary to introduce a cutoff parameter
$\Lambda$. This may be done, e.g., by restricting Fourier components of the velocity field
$\varphi$ to wave numbers less than $\Lambda$ in functional (\ref{action3}),
which automatically brings about the corresponding sharp wave-vector cutoff
in the bare response function (\ref{lines}) and in the bare correlation function (\ref{lines2}).
It was already explained in Sec. \ref{sec:local} that all such $\Lambda$ divergences are "nearly
logarithmic" and appear in the results in the form of powers $\Lambda^\alpha$ with small
(of the order of  $\eps$ for $\Delta/\eps=const$) positive exponents $\alpha$.

The elimination of the $\Lambda$ divergences may be reduced to a renormalization of the bare
parameters. Denoting for brevity the whole set of parameters by $e$ we introduce the notion of
"primary bare parameters"
$\tilde{e}_0 = \{\tilde{\nu}_0,\widetilde{D}_{i0} = \tilde{g}_{i0} \tilde{\nu}_0^3, \, i=1,2\} $
and "secondary bare parameters" $e_0= \{\nu_0,D_{i0} = g_{i0} \nu_0^3, \,i=1,2\}$ (see Ref. \cite{Kniga}).
The original model is defined by a functional of the type of Eq. (\ref{action3}) with the
$\Lambda$ cutoff introduced and with the "primary bare parameters" $\tilde{e}_0$:
\begin{equation}
%\nonumber
 S(\Phi )=\varphi '(\widetilde{D}_{1 0}k^{2-2\Delta-2\eps}+\widetilde{D}_{2 0}k^{2} )\varphi '/2
 +\varphi '[-\partial_t\varphi +\tilde{\nu} _0\partial^{2} \varphi -(\varphi
\partial )\varphi ] \,. \label{action4}
\end{equation}
Renormalization of this model may be carried out in two steps: the first is the $\Lambda$ renormalization
with the aim of removal of all $\Lambda$ divergences. This amounts to a reorganization of the
bare parameters $\tilde{e}_0\to e_0$, in which the secondary set of parameters is expressed as
functions of the primary set: $e_0=e_0(\tilde{e}_0 ,\Lambda)$;
and vice versa $\tilde{e}_0 = \tilde{e}_0  (e_0,\Lambda)$. The correspondence between the two sets
$\tilde{e}_0$ and $e_0$ is bijective perturbatively, therefore any of them may be chosen as the
set of independent variables.

If the parameters $e_0$ are chosen as independent, then in the Green functions $\Gamma$ of
model (\ref{action4}) expressed in terms of $e_0$ and $\Lambda$ there will be no $\Lambda$ divergences left
(they all will be concentrated in the formulas connecting $\tilde{e}_0$ and $e_0$) and the limit
$\Lambda\to \infty$ may be taken in them with the result of eliminating the cutoff parameter $\Lambda$
completely from the theory. Trace of the UV divergences which brought about the positive powers of
$\Lambda$ remains, however, in the form of singularities in $\eps$ in the $\Lambda$-renormalized
quantities. This happens because in the $\Lambda$ renormalization only terms strictly growing as powers
of $\Lambda$ are removed and collected in the renormalization constants. These terms contain singularities in
$\eps$ and $\Delta$, although the unrenormalized quantities with fixed $\Lambda$ were regular functions of
$\eps$ and $\Delta$.
Consequently, in the $\Lambda$-renormalized quantities there must be terms left which are singular in
$\eps$ and $\Delta$, but remain
finite in the limit $\Lambda\to \infty$. Thus, the $\Lambda$ renormalization is a way to trade UV divergences
in the form of positive powers of the UV cutoff $\Lambda$ for poles in $\eps$, $\Delta$ and their
linear combinations in such a way that in the
$\Lambda$-renormalized quantities the limit $\Lambda\to\infty$ may be taken.

The basic conjecture is that the results obtained in this manner for the graphs of the Green functions
$\Gamma(e_0, \Lambda=\infty,...)$ (the ellipsis stands for the rest of the arguments like frequencies and
wave vectors) are exactly the same as those obtained in the "formal scheme", i.e. by analytic continuation
of all integrals without $\Lambda$ divergences on the parameter $\Delta$ from the region of
small $\Delta<0$ (more accurately $-2\eps<\Delta<0$). In this scheme, the unrenormalized action is
functional (\ref{action3}). Such an analytic continuation might be carried out without any reference to
the model regularized with the explicit wave-number cutoff $\Lambda$, which is common practice in field
theories of particle physics. There, however, it is the renormalized parameters which are the physical ones and
their bare counterparts together with the UV cutoff unphysical auxiliary quantities. In our case unrenormalized
parameters are the physical ones and therefore it is important, in principle, to keep track on their
relation to the (auxiliary) renormalized parameters, because the fixed-point values of the latter remain in
the asymptotic expressions for various correlation functions and the like.

The next step after the $\Lambda$ renormalization is the $\eps$ renormalization with the goal
of removal from all Green functions
$\Gamma(e_0,\Lambda=\infty,...)$ poles in $\eps$ for $\Delta/\eps=const$. It is carried out by
the transition from the "secondary bare parameters" $e_0$ (the same notation was used in
Sec. \ref{sec:local}) to the renormalized parameters $e =\{\nu,g_1,g_2\}$
according to relations (\ref{ZZ}).

The procedure of the $\eps$ renormalization was discussed thoroughly in Sec. \ref{sec:local}.
Let us now explain in more detail the procedure of the $\Lambda$ renormalization:
the transition from the primary bare parameters
$\tilde{e}_0$ to the secondary bare parameters $e_0$. We emphasize that at this stage we are interested
in the $\Lambda$ divergences only and regard $\eps$ and $\Delta$ as fixed parameters without any
investigation of singularities in these parameters tending to zero. We shall consider the parameters
$\tilde{e}_0$ in the graphs of the functions $\Gamma$ of model (\ref{action4}) expressed in terms
of $e_0$ and $\Lambda$ through the renormalization relations
\begin{align}
\label{ZZL}
\widetilde{ D}_{10}&= \tilde{g}_{10}\tilde{\nu}_0^{3} = g_{10} \nu_0^{3}=D_{10}\,,
\nonumber\\
\widetilde{D}_{20} &= \tilde{g}_{20}\tilde{\nu}_0^{3} = g_{20}
\nu_0^{3} {\widetilde{Z}}_{D_2}= D_{20}{\widetilde{Z}}_{D_2}\,,
\nonumber\\
\tilde{g}_{10}&=g_{10}{\widetilde{Z}}_{g_1},\qquad\tilde{g}_{20}=g_{20}{\widetilde{Z}}_{g_2}
\,,\\
\tilde{\nu}_0&={\nu_0} {\widetilde{Z}}_{\nu}\,,\quad
{\widetilde{Z}}_{g_1}{\widetilde{Z}}_{\nu}^{3}=1,\quad
{\widetilde{Z}}_{g_2}{\widetilde{Z}}_{\nu}^{3}={\widetilde{Z}}_{D_2}\,.&\nonumber
\end{align}
similar to Eq. (\ref{ZZ}). The dimensionless renormalization constants $\widetilde{Z}$ in
Eq. (\ref{ZZL}) are functions of $e_0$ and $\Lambda$ expressed in the form of series in
${D}_{i0} \sim {g}_{i0}$. The corresponding dimensionless expansion parameters are
the following analogs of Eq. (\ref{alpha}):
\begin{equation}
\tilde{\alpha}_1\equiv\frac{D_{10}\overline{S}_d}{32\nu_0^3{\Lambda}^{2\eps}}\,
,\qquad
\tilde{\alpha}_2\equiv\frac{D_{20}\overline{S}_d}{32\nu_0^3{\Lambda}^{-2\Delta}}\;\;.
\label{alpha3}
\end{equation}
Therefore, the constants $\widetilde{Z}$ in Eq. (\ref{ZZL}) assume the
form
\be
 \widetilde{Z}_{\nu,D_2}=1+\hskip -0.5cm\sum_{{\,{\,n}_1}\ge0,{\,{\,n}_2}\ge1}\hskip -0.4cm
 C_{\nu,D_2}^{({{\,n}_1},{\,{\,n}_2})}\;\tilde{\alpha}_1^{\,{\,n}_1}\tilde{\alpha}_2^{\,{\,n}_2}
\label{ZL}
\ee
with the dimensionless coefficients $C_{\nu,D_2}^{({{\,n}_1},{\,{\,n}_2})}$ depending on
$\eps$  and $\Delta/\eps=\zeta$ only (but in a singular manner!).
In expansion (\ref{ZL}) not all possible terms are included,
but only those which are "$\Lambda$ divergent", i.e. those with a positive power of $\Lambda$
in the product $\tilde{\alpha}_1^{n_1} \tilde{\alpha}_2^{n_2}$. From Eq. (\ref{alpha3}) it follows
\be
       \tilde{\alpha}_1^{n_1} \tilde{\alpha}_2^{n_2} \sim
       \Lambda^\alpha, \quad
       \alpha = 2(n_2 \Delta-n_1 \eps),
\ee
therefore, for $\eps>0,\Delta>0$, in the $\Lambda$-divergent terms with $\alpha>0$ in Eq. (\ref{ZL})
the inequality $n_2\ge 1$ holds, i.e. at least one factor
with $\widetilde{D}_{20}\sim D_{20}$ from (\ref{action4}) is present.

From this it follows, in particular, that
to the real value $\widetilde{D}_{20}=0$ in Eq. (\ref{action4}) [see the text after Eq. (\ref{nakach2})]
it corresponds $D_{20}=0$ in Eq. (\ref{action3}), which justifies the derivation of Eq.
(\ref{2.74}) from Eq. (\ref{bal}) in model (\ref{action3}). We also note that
the operation $\Dm$ in Eq. (\ref{gnu}), defined in Sec. \ref{sec:RGE} as  $\Dm\equiv \mu\partial_{\mu}$
with fixed parameters $e_0$, in terms of model (\ref{action4}) has to be
understood as $\mu\partial_{\mu}$ with fixed $\widetilde{e}_0$ and $\Lambda$. These definitions
are equivalent, because the parameter $\mu$ does not enter in renormalization relations (\ref{ZZL}).

For the $\Lambda$ renormalization (\ref{ZZL}) analogs of relations (\ref{ti3})-(\ref{Rat2}) may be
written and the corresponding $\widetilde{Z}$ calculated at two-loop order. We shall not quote the
corresponding results, because explicit expressions connecting the primary
($\widetilde{e}_0$) and secondary ($e_0$) bare parameters is unimportant for the RG analysis
of the IR asymptotic behavior in Sec. \ref{sec:RGE}, which is carried out in terms of
bare parameters $e_0$ and renormalized parameters $e$.

In the NP scheme the normalization condition (\ref{Rato3}) may be imposed
in the $\Lambda$-renormalized model in the same way as just described for the MS (or $\overline{\text{MS}}$) scheme.
It is not difficult to see, however, that in the NP scheme the very procedure of the $\Lambda$ renormalization
is actually not necessary. The point is that in this scheme from the quantity to be renormalized its value at
the normalization point is subtracted which automatically leads to a quantity without any UV divergences and
thus with a finite -- and regular in $\eps$ and $\Delta$ -- limit, when $\Lambda\to \infty$. For renormalized
correlation functions the result is the same as after $\Lambda$ renormalization, subsequent limit
$\Lambda\to\infty$ and final renormalization in the NP scheme. Therefore, the RG functions $\gamma$ and
$\beta$ are also the same, since their expressions in terms of the renormalized correlation functions
coincide in both cases.

\end{document}